\ifx\documentclass\undefined
\documentstyle[12pt]{article}
\else
\documentclass[12pt]{article}
\fi

\usepackage{color}

\usepackage{ascmac}
\usepackage{amsmath}
\usepackage{amssymb}

\usepackage{here}

\usepackage[dvipdfmx]{graphicx}
\usepackage{pdfpages}
\usepackage{amsmath}

\newcommand{\beq}{\begin{eqnarray*}}
\newcommand{\eeq}{\end{eqnarray*}}
\makeatletter
\renewcommand{\theequation}{\thesection.\arabic{equation}}
\@addtoreset{equation}{section}
\def\eqnarray{%
\stepcounter{equation}%
\let\@currentlabel=\theequation
\global\@eqnswtrue
\global\@eqcnt\z@
\tabskip\@centering
\let\\=\@eqncr
$$\halign to \displaywidth\bgroup\@eqnsel\hskip\@centering
$\displaystyle\tabskip\z@{##}$&\global\@eqcnt\@ne
\hfil$\displaystyle{{}##{}}$\hfil
&\global\@eqcnt\tw@$\displaystyle\tabskip\z@{##}$\hfil
\tabskip\@centering&\llap{##}\tabskip\z@\cr}
\makeatother
\newsavebox{\toy}
\savebox{\toy}{\framebox[0.65em]{\rule{0cm}{1ex}}}
\newcommand{\QED}{\usebox{\toy}}
\def\nlni{\par\ifvmode\removelastskip\fi\vskip\baselineskip\noindent}
\newenvironment{proof}{\nlni\begingroup\it Proof.\rm}{
\endgroup\vskip\baselineskip}
\newcommand{\supp}{\mathop{\mathrm{supp}}\nolimits}
\begin{document}
%%%%%%% DOUBLE SPACED %%%%%%%%
\setlength{\baselineskip}{15pt}
\title{
Eigenvalue Fluctuations of 
1-dimensional random Schr\"odinger operators
}
\author{
Takuto Mashiko, 
Yuma Marui, 
Naoki Maruyama, 
and 
Fumihiko Nakano
\thanks{
Mathematical Institute,
Tohoku University,
Sendai 980-8578, Japan
e-mail : 
fumihiko.nakano.e4@tohoku.ac.jp
}
}
%\date{最終更新日：}
\maketitle
%第一ページの番号を消す
%\thispagestyle{empty}
%%%%%%% ABSTRACT %%%%%%%%%%%%%
\begin{abstract}
As an extension 
to the paper by 
Breuer, Grinshpon, and White
\cite{B}, 
we study
the linear statistics for the eigenvalues of the Schr\"odinger operator with random decaying potential with order 
${\cal O}(x^{-\alpha})$
($\alpha>0$)
at infinity. 
We first prove 
similar statements as in 
\cite{B} 
for the trace of 
$f(H)$, 
where 
$f$ 
belongs to a class of analytic functions : 
there exists a critical exponent 
$\alpha_c$
such that 
the fluctuation of the trace of 
$f(H)$ 
converges in probability for 
$\alpha > \alpha_c$, 
and satisfies a CLT statement for 
$\alpha \le \alpha_c$, 
where 
$\alpha_c$
differs depending on 
$f$. 
Furthermore 
we study the asymptotic behavior of its expectation value. 
\end{abstract}

%Mathematics Subject Classification (2000): 82B44, 81Q10

%\tableofcontents
%%%%% INTRODUCTION %%%%%%%%%%%%%%%%%%%%%%%%%%%%%%%%%
\section{Introduction}
The study 
of the one-dimensional Schr\"odinger operator with random decaying potential was initiated by 
\cite{DKS}
where they found that it has various spectral properties depending on the decay exponent.
After the discovery 
\cite{DE}
of the connection between the Jacobi matrices and beta-ensemble, there appear many papers studying from the RMT-point of view, e.g., 
the eigenvalue statistics on the bulk
\cite{KVV, KN1, N1, KN2},
linear statistics 
\cite{N2, B}, 
and eigenfunction statistics 
\cite{RV, N3}. 
In this paper, 
we consider the following Hamiltonian, 
to extend the work by 
Breuer, Grinshpon, and White
\cite{B}.
\beq
(H_{\alpha} u)_n
&:=&
\begin{cases}
u_{n+1} + u_{n-1} + V_{\alpha} (n) u_n & (n \ge 2) \\
u_2 + V_{\alpha} (1) u_1 & (n=1) 
\end{cases}
\\
V_{\alpha} (n)
&:=&
\frac {X_n}{n^{\alpha}}, 
\quad
n \in {\bf N}
\eeq
where 
$\{ X_n \}$
is a family of i.i.d. random varables.
For 
$N \in {\bf N}$, 
let 
$H_{\alpha, N}$
be the restriction of 
$H_{\alpha}$ 
on 
$[1, N]$.
Breuer, Grinshpon, and White
\cite{B}
studied the fluctuation of 
$Tr (P(H_{\alpha, N}))$ 
and showed that 
we can find a critical exponent 
$\alpha_c$
depending on 
$P \in {\cal P}_m$
such that the behavior of the fluctuation of 
$Tr P(H_{\alpha, N})$
changes drastically at 
$\alpha_c$ : 
\beq
&(1)& \quad
(\alpha > \alpha_c)
\quad
Tr P (H_{\alpha, N})
-
{\bf E}[
Tr P (H_{\alpha, N})
]
\stackrel{a.s.}{\to} 
Q
\\
&(2)& \quad
(\alpha \le \alpha_c)
\quad
\frac {1}{
g_{\alpha/\alpha_c}(N)
}
\left\{
Tr P (H_{\alpha, N})
-
{\bf E}[
Tr P (H_{\alpha, N})
]
\right\}
\stackrel{d}{\to}
N(0, \sigma (P)^2)
\eeq
where 
$Q$
is a random variable with finite variance, and 
\beq
g_{t}(N)
&:=&
\begin{cases}
\dfrac {N^{1 - t}}{1 - t}
&
(0 < t < 1)
\\
\log N
%\frac {N^{1 - 2 \alpha}}{1 - 2 \alpha}
&
(t = 1)
\end{cases}
\eeq
$\sigma (P) \ge 0$
is an explicit constant.
Moreover, 
the space 
${\cal P}_m$ 
of polynomials of degree 
$m$
has a decomposition 
${\cal P}_m
=
{\cal Q}_m \oplus {\cal E}_m \oplus {\cal Q}_m^{\perp}$
such that 
\beq
\alpha_c = 
\begin{cases}
1/2 & (P \in {\cal Q}_m) \\
1/4 & (P \in {\cal E}_m) \\
1/6 & (P \in {\cal Q}^{\perp}_m) \\
\end{cases}
\eeq
On the other hand, 
\cite{N2}
studied its continuum analogue of 
$H_{\alpha}$ 
and derived the asymptotics of 
$Tr (1_{(a,b)}(H_{\alpha, N})$
($0 < a < b$).
\beq
&(1)& 
\quad
(\alpha > 1/2)
\quad
Tr (1_{(a,b)}(H_{\alpha, N})
-
N \mu(a,b)
\stackrel{a.s.}{\to}
Q
\\
&(2)& 
\quad
(\alpha = 1/2)
\quad
(\log N)^{-1/2}
\left\{
Tr (1_{(a,b)}(H_{\alpha, N}))
-
N \mu(a,b)
-
(Const.)
\log N
\right\}
\stackrel{a.s.}{\to}
G(a,b)
\\
&(3)& 
\quad
(\alpha < 1/2)
\quad
N^{-(1/2 - \alpha)}
\left\{
Tr (1_{(a,b)}(H_{\alpha, N}))
-
N \mu(a,b)
-
\sum_{k=1}^{D+1}
C_k
N^{1 - k \alpha}
\right\}
\stackrel{a.s.}{\to}
G(a,b)
\eeq
where 
$\mu$ 
is the IDS
(in the continuum, 
$\mu(a,b) 
=
\pi^{-1}(\sqrt{b} - \sqrt{a})$).
$Q$
is a bounded random variable.
$D := 
\min
\{ 
d \in {\bf N}
\, | \, 
\frac {1}{2 \alpha} < d + 1 
\}$, 
and 
$\{ G(a,b) \}_{0 < a < b}$, 
is a Gaussian field.
(2), (3)
holds in the sense of joint distribution w.r.t. 
$a, b$.
In view of 
\cite{B, N2}, 
we consider the following problems : 
(1)
we consider 
$Tr f(H_{\alpha, N})$
instead of 
$Tr P(H_{\alpha, N})$
where 
$f$
is an analytic function.
(2)
For a number of 
analytic functions 
$f_1, f_2, \cdots, f_K$, 
we consider the joint limit of 
$Tr f_j (H_{\alpha, N})
-
{\bf E}[
Tr f_j (H_{\alpha, N})
]$, 
$(j=1, 2, \cdots, K)$, 
and 
(3)
We study 
the behavior of 
${\bf E}[ f (H_{\alpha, N}) ] $.
We possibly 
would like to 
$Tr 1_{(a,b)}(H_{\alpha, N})$
but for that we need to study Pr\"ufer coordinate 
\cite{KLS}
which is beyond the scope of this paper. 
It 
would be difficult to consider general continuous function, so that we work 
under the following condition in this paper.\\

{\bf Assumption }\\
{\it 
(1)
$\{ X_n \}$
is a family of i.i.d. bounded random varables, such that 
${\bf E}[ X_n ] = 0$, 
$\eta^2 := {\bf E}[ X_n^2 ] > 0$, 
and 
$| X_n (\omega) | \le C_X$, a.s.
for a positive constant 
$C_X$.\\
(2)
$f$
has the Taylor expansion around the origin and its convergence radius satisfies 
$r(f) > C_X + 2$.
\beq
f (x)
= 
\sum_{j=0}^{\infty}
c_j x^j, 
\quad
|x| < r(f), 
\quad
r(f) > C_X + 2.
\eeq
}
Since 
$\sigma (H_{\alpha})
\subset
[-2- C_X, 2 + C_X]$, 
the assumption on 
$r(f)$ 
is a natural one ensuring the convergence of 
$f(H) = 
\sum_j c_j H^j$
in the norm topology.
We state 
our results in each subsections below. 
%

%
%%%%%%%%%%%%%%%%%%%%%%%%%%%%%%%%%%%%
\subsection{Analytic extension}
Let 
$f_k$, 
$k=1, 2, \cdots$
be the polynomial obtained by truncating higher order terms in the Taylor expansion of $f$ : 
\beq
f_k(x)
&=&
\sum_{ j = 0}^k c_j x^j. 
\eeq
And let 
\begin{equation}
p^k (\beta)
:=
\sharp 
{\cal P}^k (\beta)
\label{p}
\end{equation}
be the number paths where 
the initial and terminal points coincide and the number of flat steps on each site is given by a multi-index  
$\beta$. 
The precise 
definition is given in 
(\ref{pprecise}).

Then 
we have an analogues statement as in 
\cite{B} : 
we can find a critical exponent 
$\alpha_{c}$ 
such that for 
$\alpha_{c} < \alpha$ 
the fluctuation converges to a random variable, while for 
$0 < \alpha \le \alpha_{c}$ 
the fluctuation is unbounded and satisfies a CLT statement after a suitable scaling.\\
{\bf Theorem 1}\\
%
%%%% italic begins %%%%%
{\it 
{\bf Case A :} general case \\
(1)
$1/2 < \alpha$ : 
there exists a random variable 
$Q_A$ 
with 
$var (Q_A) < \infty$ 
s.t. 
\beq
Tr f(H_{\alpha, N}) - {\bf E}[ Tr f(H_{\alpha, N}) ]
\stackrel{N \to \infty}{\to}
Q_A, 
\quad
\mbox{ a.s. }
\eeq
If 
$\alpha > 1$, 
$Q_A$
is bounded. \\
(2)
$0 < \alpha \le 1/2$ : 
we have a following CLT statement. 
\beq
\frac {
Tr f(H_{\alpha, N}) - {\bf E}[ Tr f(H_{\alpha, N}) ]}
{
g_{2 \alpha} (N)
}
\stackrel{d}{\to}
N(0, \sigma_A(f)^2),
\eeq
where 
\beq
\sigma_A (f)^2
:=
\left|
\sum_{j=0}^{\infty}
c_{2j+1} p^{2j+1} (\delta)
\right|^2
\eta^2, 
\quad
p^k (\delta)
=
k 
\binom {k-1}{\frac {k-1}{2}}
1(k : \mbox{odd}).
\eeq
%

%%%%%
%

%
%%%%% B %%%%%%%%%%%
{\bf Case B :}
Taylor expansion of 
$f$
have even terms only : 
$c_{2j-1} = 0$, $j \in {\bf N}$.
\\
(1)
$1/4 < \alpha$ : 
there exists a random variable 
$Q_B$ 
with 
$var (Q_B) < \infty$ 
s.t. 
\beq
Tr f(H_{\alpha, N}) - {\bf E}[ Tr f(H_{\alpha, N}) ]
\stackrel{N \to \infty}{\to}
Q_B
\quad
\mbox{ in probability. }
\eeq
(2)
$0 < \alpha \le 1/4$ : 
\beq
&&
\frac {
Tr f(H_{\alpha, N})
-
{\bf E}[ Tr f(H_{\alpha, N}) ]
}
{
g_{4 \alpha} (N)
}
\stackrel{d}{\to}
N(0, \sigma_{B} (f)^2 )
\eeq
where 
\beq
%\sigma_{\cal E} (f)^2
%&:=&
%\lim_{k \to \infty}
%\sigma_{\cal E} (f_k)^2
%\\
%
\sigma_{B} (f)^2
&:=&
\left(
\sum_{j =1}^{\infty}
c_j p^j (2\delta)
\right)^2
\left(
{\bf E}[ X_1^4 ] - \eta^4 
\right)
+
\sum_{s=1}^{\infty}
\left(
\sum_{j=1}^{\infty}
c_j
p^j (\delta + \delta^s)
\right)^2
\eta^4.
\eeq
{\bf Case C :} 
Taylor 
expansion of 
$f$
have odd terms only : 
$c_{2j} = 0$, 
$j \in {\bf N}$, 
and 
$f$
has the form of 
\beq
f(x)
&=&
\sum_{j=3}^{\infty}
c_j
\left(
x^j - p^j (\delta) x
\right)
=
\sum_{j=3}^{\infty}
c_j x^j
-
\left(
\sum_{j=3}^{\infty} 
p^j (\delta)
\right) 
x.
\eeq
(1)
$1/6 < \alpha$ : 
there exists a random variable 
$Q_C$ 
with 
$var (Q_C) < \infty$ 
s.t. 
\beq
Tr f(H_{\alpha, N}) - {\bf E}[ Tr f(H_{\alpha, N}) ]
\stackrel{N \to \infty}{\to}
Q_C
\quad
\mbox{ in probability. }
\eeq
(2)
$0 < \alpha \le 1/6$ : 
\beq
&&
\frac {
Tr f(H_{\alpha, N})
-
{\bf E}[ Tr f(H_{\alpha, N}) ]
}
{
g_{6 \alpha} (N)
}
\stackrel{d}{\to}
N(0, \sigma_{C} (f)^2 )
\eeq
where
\beq
%\mbox{ where }
%\;
\sigma_{C} (f)^2
&:=&
\lim_{k \to \infty}
\sigma_{C} (f_k)^2.
\eeq
}
%%%% italic is over %%%%%

%
The expression of 
$\sigma_{C} (f_k)$, 
given in \cite{B}(2.11), 
is so complicated and is omitted. 
We believe 
that the convergence in Theorem 1(1) in cases B and C
%B(1), C(1)
also holds almost surely, not only in probability. 
In fact, 
we can show the a.s. convergence if we interchange the order of summation in a random series (Lemma 2.9).

%
%%%%%%%%%%%%%%%%%%%%%%%%%%%%%%%%%%%%
\subsection{Joint limit}
Suppose 
$f_t$
($t=1, 2, \cdots, d$)
satisfy Assumption.
We assume that 
they belong to same cases among A, B, and C, and set 
$\alpha_c := 
1/2$ (Case A), 
$1/4$ (Case B), 
and 
$1/6$ (Case C),
and similarly, 
$\sigma (f) := 
\sigma_{\sharp} (f)$
for Case
$\sharp$ 
($\sharp = A, B, C$).
We consider the joint limit of the following vector : 
\beq
{\bf F}_N 
&:=&
(
F_{1, N}, \cdots, F_{d, N}
)
\\
\mbox{ where }
\quad
F_{t, N}
&:=&
\frac {
Tr f_t(H_{\alpha, N}) - {\bf E}[ Tr f_t(H_{\alpha, N}) ]}
{
g_{\alpha/\alpha_c} (N)
}, 
\quad
t = 1, 2, \cdots, d.
\eeq
%

%
%%%%%%%%%
{\bf Theorem 2 }
{\it 
Let 
$0 < \alpha \le \alpha_c$
}
\beq
{\bf F}_N
&\stackrel{d}{\to}&
{\bf F}, 
\quad
where
\quad
{\bf F}
\sim
N(0, \Sigma_{\bf F}), 
\quad
\Sigma_{\bf F}
:=
\left(
\sigma (f_i) \sigma (f_j)
\right)_{i, j}.
\eeq
%

%
%%%%%%%%%%%%%%%%%%%%%%%%%%%%%%%%%%%%%%%%%%%%%%%%%%%%%%%%%%%%
\subsection{Behavior 
of the expectation value}
Let 
$m :=
\max \{ k \in {\bf N}
\, | \,
k \alpha \le 1 \}$.
Then 
we have the asymptotic expansion for the expectation value of the trace of 
$f(H)$.\\
{\bf Theorem 3}
{\it 
\beq
&&
{\bf E}
[ Tr f(H_{\alpha, N}) ] 
\\
&\stackrel{}{=}&
\left(
\sum_l
c_l C_{0,l}
\right)
N
+
\left(
\sum_l
c_l C_{2,l}
\right)
S_2 (N)
%N^{1 - 2 \alpha}
+
\cdots
+
\left(
\sum_l
c_l C_{m,l}
\right)
S_m (N)
%N^{1 - m \alpha}
+
C_N(f)
%（有界項）
\eeq
where 
\beq
S_j
(N)
:=
\sum_{n=1}^N
\frac {1}{
n^{j \alpha}
}
=
\begin{cases}
\dfrac {
N^{1 - j \alpha}
}
{
1 - j \alpha
}
+ {\cal O}(1)
%(1 + o(1))
&
(j \alpha <1) 
\\
\log N
+ {\cal O}(1)
%(1 + o(1))
&
(j \alpha = 1)
\end{cases}
%+{\cal O}(1), 
\quad
j = 1, 2, \cdots, m
\eeq
$C_{0, l}, \cdots, C_{m, l}$
are explicit constants, and 
$C_N (f)$
is convergent as 
$N \to \infty$.
}\\

In later sections, 
we prove these theorems.
In Section 2, 
we first recall the argument in 
\cite{B}
which relates 
$Tr (H^k)$ 
to the path counting on 
${\bf Z}$.
Then 
we prepare some lemmas to prove Theorem 1.
In Section 3, 
we prove Theorem 1.
Proof of  
the CLT statement is a simple application of a limit theorem on the triangular array of random variables. 
However, 
to show Theorem 1(1) in Cases B and C, 
%B(1), C(1)
one needs complicated computations and estimates of the variance. 
In Section 4, 
we prove Theorem 2 using Cramer-Wald's method. 
In Section 5 (resp. in Section 6), 
we prove Theorem 3
for 
$f(x) = x^k$ 
(resp. for analytic functions). 
%

%%%%%%%%%%%%%%%%%%%%%%%%%%%%%%%%%%%%%%%%%%%%%%%%%%%%%%%%%%%%
\section{Preliminaries}
%
%%%%%%%%%%
\subsection{Paths with flat steps}
Since 
$Tr H^k_{\alpha, N}$
is a polynomial of 
$V(1), \cdots, V(N)$, 
it is important to derive the coefficients of that.
In 
\cite{B}, 
they express them in terms of the lattice path on 
${\bf Z}$
with certain conditions.
We 
briefly recall this argument in 
\cite{B}, 
not only for completeness but also to fix the notation.\\
{\bf Definition 2.1}\\
{\it 
(1)
A multi-index
is a map 
$\beta : {\bf Z} \to {\bf N} \cup \{ 0 \}$
such that 
$\beta (h) = 0$
but finitely many 
$h$'s.
We sometimes write 
$\beta_h$
instead of 
$\beta(h)$.
For a multi-index
$\beta$, 
let 
$V^{\beta}$ 
be the monimial given by 
\beq
V^{\beta}
:=
\prod_{h \in {\bf Z}}
V(h)^{\beta(h)}.
\eeq
We set 
$V(h) = 0$
for 
$h \le 0$, 
and also set 
$0^n := 1$
for convenience. 
The multiplicity of 
$V^{\beta}$ 
is denoted by 
$|\beta| := 
\sum_{j=1}^N \beta(j)$.
The translation 
$\beta^i$ 
of 
$\beta$ 
for 
$i \in {\bf Z}$
is defined by 
\beq
(\beta^i) (h)
:=
\beta(h - i), 
\quad
h \in {\bf Z}.
\eeq
(2)
$\delta = \delta_0$
is a multi-index with 
\beq
\delta(h)
:=
1(h = 0).
\eeq
%
%

%%%
%

%
%%%%%%%%%
{\bf Definition 2.2}\\
(1)
Path on 
${\bf Z}$
of length 
$k$
is a sequence 
${\bf y}:=(y_0, y_1, \cdots, y_k)$
$(y_i \in {\bf Z})$
such that 
$| y_t - y_{t-1} | \le 1$.
We say 
a pair 
$(y_{t-1}, y_t)$
is a up step (resp. down step, flat step) 
if  
$y_t - y_{t-1}=1$
(resp. $y_t - y_{t-1}=-1$, 
$y_t - y_{t-1}=0$).
We say 
$(y_{t-1}, y_t)$ 
is a flat step of level 
$h \in {\bf Z}$ 
if 
$y_{t-1} = y_t = h \in {\bf Z}$.
Let 
${\cal Q}^k$
be the set of paths of length 
$k$
starting from the origin : 
\beq
{\cal Q}^k
&:=&
\left\{
{\bf y}
=
(y_0, y_1, \cdots, y_k)
\, \middle| \,
y_t \in {\bf Z}, 
\;
y_t - y_{t-1}= \pm 1, 0 
\;
(t=1, 2, \cdots, k), 
\;
y_0 = 0
\right\}.
\eeq
For a given multi-index
$\beta$, 
let 
${\cal Q}^k (\beta)$
be the set of paths in 
${\cal Q}^k$
such that the number of flat steps of which is counted by a suitable translation of 
$\beta$ : 
\beq
{\cal Q}^k 
(\beta)
&:=&
\left\{
{\bf y}
=
(y_0, y_1, \cdots, y_k)
\in {\cal Q}^k
\, \middle| \,
\exists i \in {\bf Z}
\;s.t.\;
\beta^i(h)
=
\sharp\{
t
\, | \,
y_{t-1} = y_t = h
\}, 
\;
\forall h
\right\}
\eeq
We note that, for 
${\bf y} \in {\cal Q}^k (\beta)$, 
the corresponding 
$i \in {\bf Z}$
is uniquely determined. 
\\
}

We aim to relate 
each terms in 
$Tr (H^k_{\alpha, N})$
(as a polynomial of 
$V$)
to the paths in 
${\cal Q}^k$.
In order for that, 
we write 
\beq
H_{\alpha, N}
&=&
S_N + V_{\alpha, N} + S_N^*
\eeq
where 
$S_N$ 
is the right shift on 
$\ell^2 ([1, N])$.
Expanding 
$(S_N + V_{\alpha, N} + S_N^*)^k$ 
and taking each terms, we have 
\beq
H_{\alpha, N}^k
&=&
(S_N + V_{\alpha, N} + S_N^*)^k
=
\sum_{M \in {\cal M}} M
\eeq
where 
${\cal M}$
is the set of strings of 
$S_N, S_N^*, V_{\alpha, N}$
of length 
$k$ : 
\beq
{\cal M}
&:=&
\left\{
%\prod_{t=1}^k M_t
M_1 M_2 \cdots M_k
\, \middle| \,
M_t = S_N, S_N^*, V_{\alpha, N}, 
\;
t=1, 2, \cdots, k
\right\}.
\eeq
For 
$M = 
%\prod_{i=1}^k M_i 
M_1 M_2 \cdots M_k\in {\cal M}$, 
we set the corresponding path 
${\bf y}
=
(y_0, y_1, \cdots, y_k) \in {\cal Q}^k$
as 
\beq
y_0 := 0, 
\quad
y_t - y_{t-1}
&:=&
\begin{cases}
1 & (M_t = S_N) \\
-1 & (M_t = S_N^*) \\
0 & (M_t = V_{\alpha, N}) 
\end{cases}
\eeq
which gives a bijection between 
${\cal M}$
and 
${\cal P}^k$.
We henceforth write 
${\bf y}(M) \in {\cal Q}^k$
is a path corresponding to 
$M \in {\cal M}$
and vice versa.\\
%

%
%%%%%
{\bf Lemma 2.1 }\\
{\it 
For 
$M = 
%\prod_{t=1}^k M_t 
M_1 M_2 \cdots M_k\in {\cal M}$, 
let 
${\bf y}(M)=(y_0, y_1, \cdots, y_k) \in {\cal Q}^k$
be the corresponding path.
For 
$i, j \in [N]$, 
we have 
\beq
M_{ij}
&=&
V^{\beta}
\cdot
1
\left(
\beta, 
{\bf y}(M)
\mbox{ 
satisfies condition 
}
(C)_{i, j}^{\beta}
\right)
\eeq
where 
$1(P)$
is the indicator function associated to a proposition 
$P$ 
and we say a multi-index
$\beta$ 
and 
${\bf y} \in {\cal P}^k$
satisfies condition 
$(C)_{i, j}^{\beta}$
if \\
($C_1$)
${\bf y} + i$
lies in 
$[1, N]$ : 
\beq
1 \le i + y_t \le N, 
\quad
1 \le t \le k. 
\eeq
($C_2$)
$\beta^{-i}$
counts the number of flat steps of 
${\bf y}$ : 
\beq
\sharp \{
t 
\, | \,
y_t = y_{t-1} = h 
\}
=
\beta(h+i), 
\quad
\forall h \in {\bf Z}. 
\eeq
($C_3$)
$j - i = y_k (M)$. 
}
\\
Note that,  
the condition 
$(C_2)$
determines 
$\beta$
uniquely, 
and 
for each
$i \in [N]$, 
there is at most one 
$j$
such that the condition
$(C)_{i, j}^{\beta}$
is satisfied.
%

%%%%%
%
\begin{proof}
For simplicity of notation, 
we write 
$U := S_N$, 
$D := S_N^*$.
On the products in 
$M = 
%\prod_{t=1}^k M_t 
M_1 M_2 \cdots M_k\in {\cal M}$,
we group together the products of 
$U$'s 
which are divided by an appearance of 
$V$, 
and denote by 
$W_1, W_2, \cdots$ 
the products of 
$U$'s 
in each group : 
\beq
M
&=&
M_1 M_2 \cdots M_k
%\prod_{t} M_t
=
(U D \cdots U)
V
(D U \cdots D)
V
\cdots
V 
(D U \cdots U)
\\
&=:&
W_1 V W_2 V \cdots W_{f} V W_{f+1}
\eeq
where the number of flat steps is equal to 
$f$ 
in 
${\bf y}(M)$. 
For each 
$W_l$
($l = 1, 2, \cdots, f+1$), 
let 
$s_l
:=
\sharp U - \sharp D$
be the differeces between the number of 
$U$'s  
and that of 
$V$'s.
Letting 
$y_t (n_l \le t \le m_l)$
be the part of
${\bf y} (M)$ 
associated to 
$W_l$, 
we have 
\beq
(W_l)_{ij}
=
1 
\left(
1 \le i + y_t \le N, 
\;
n_l \le t \le m_l 
\right)
\cdot
1 
\left(
j = i + s_l
\right)
\eeq
so that 
\beq
M_{ij}
&=&
V(i+s_1)
V(i+s_1 + s_2)
\cdots
V(i+s_1+s_2+ \cdots + s_{f})
\times
\\
&& 
\qquad
\times
1 \left(
1 \le i + y_t \le N, 
\;
1 \le t \le k
\right)
\cdot
1 \left(
j = i + y_k 
\right)
\eeq
Because the set 
%
%\beq
$
\left\{
i + s_1, 
i + s_1 + s_2, 
\cdots,
i + s_1 + \cdots + s_f
\right\}
$
coincides as a multi-set with that of levels 
$\{ h \, |\, 
\exists t \in \{ 0, 1, \cdots, k-1 \}, 
\,
s. 
\,
t.
\,
y_{t-1} = y_t = h 
\}$
of flat steps of 
${\bf y}(M)$, 
we have 
\beq
M_{ij}
&=&
V^{\beta}
\cdot
1 \left(
1 \le i + y_t \le N, 
\;
1 \le t \le k, 
\;
j = i + y_k 
\right)
\\
where 
\quad
\beta(h+i)
&=&
\sharp \{
t 
\, | \,
y_t = y_{t-1} = h 
\}, 
\quad
\forall h \in {\bf Z}.
\eeq
\QED
\end{proof}
In what follows, 
we consider the diagonal element of 
$H_{\alpha, N}^k$
and set 
$i = j$, 
in the corresponding paths of which the number of up steps is equal to that of down steps.
We introduce 
the subset of 
${\cal Q}^k$ 
returning to the origin satisfying  
$(C)_{i, i}^{\beta}$, 
and those satisfying 
$(C_2)$
only : 
%
%
%\begin{eqnarray}
\beq
{\cal C}^k_i (\beta)
&:=&
\left\{
{\bf y} \in {\cal Q}^k
\, \middle| \,
y_0 = y_k = 0, 
\;
\beta, 
{\bf y}
\mbox{ 
satisfies }
(C)_{i, i}^{\beta}
\right\}
%\nonumber
\\
{\cal P}^k_i (\beta)
&:=&
\left\{
{\bf y} \in {\cal Q}^k
\, \middle| \,
y_0 = y_k = 0, 
\;
\sharp \{
t 
\, | \,
y_t = y_{t-1} = h 
\}
=
\beta(h+i), 
\;
\forall h \in {\bf Z}
\right\}
%\nonumber
\\
{\cal P}^k (\beta)
&:=&
\bigcup_i
{\cal P}_i^k (\beta)
%\nonumber
\eeq
%
%p^k (\beta)
%&:=&
%\sharp 
%{\cal P}^k (\beta)
%\label{p}
%\end{eqnarray}
%
By definition, 
${\cal C}^k_i (\beta) 
\subset 
{\cal P}^k_i (\beta)$ 
and 
${\cal P}^k (\beta)
=\bigcup_{i}
{\cal P}^k_i (\beta)
$
is a disjoint union.
And let 
\begin{equation}
p^k (\beta)
:=
\sharp 
{\cal P}^k (\beta)
\label{pprecise}
\end{equation}
be the number paths in 
${\cal P}^k (\beta)$, 
which is the number of paths where the number of flat steps on each site is given by a multi-index  
$\beta$. 
For a polynomial 
$P$, 
we denote by 
$[x^k] P$
the coefficient of 
$x^k$
in 
$P$.
\\

%
%%%%%
{\bf Lemma 2.2}
\beq
&(1)& \quad
[V^{\beta}] (H^k_{\alpha, N})_{ii}
=
\sharp
{\cal C}^k_i (\beta),
\\
&(2)& \quad
\frac k2 \le i \le N - \frac k2
\quad
\Longrightarrow
\quad
[ V^{\beta} ] 
(H^k_{\alpha, N})_{ii}
=
\sharp
{\cal P}^k_i (\beta).
\eeq
%

%%%%%
%
\begin{proof}
(1)
Letting 
$i = j$ 
in Lemma 2.1, 
we have 
\beq
M_{ii} = V^{\beta}
\cdot
1 \left(
{\bf y}(M)
\in 
{\cal C}^k_i (\beta)
\right). 
\eeq
We thus have
\beq
(H^k_{\alpha, N})_{ii}
&=&
\sum_{ {\bf y} \in {\cal P}^k }
M_{ii} ({\bf y})
=
\sum_{\beta}
\sum_{{\bf y} \in {\cal C}^k_i (\beta)}
M_{ii} ({\bf y})
=
\sum_{\beta}
\sum_{{\bf y} \in {\cal C}^k_i (\beta)}
V^{\beta}
=
\sum_{\beta}
\sharp
{\cal C}_i^k (\beta) 
V^{\beta}.
\eeq
(2)
If 
$\frac k2 \le i \le N - \frac k2$ 
then the condition 
$(C_1)$
is always satisfied and 
${\cal C}^k_i (\beta)
=
{\cal P}^k_i (\beta)$.
\QED
\end{proof}
For a multi-index
$\beta$, 
let
\beq
\iota(\beta)
&:=&
\min \supp \beta
\\
a^k_N (\beta)
&:=&
[V^{\beta}]
Tr (H^k_{\alpha, N}).
\eeq
%
%
%%%%%
{\bf Lemma 2.3 }
{\it 
If 
$\beta \ne 0$, 
and 
$\iota(\beta) \in [k, N-k]$, 
then 
$a^k_N (\beta)
%[V^{\beta}]Tr (H^k_{\alpha, N})
=
p^k (\beta)$.
}
%
%%%%%
%
\begin{proof}
We note that, if
$\iota (\beta) \in [k, N-k]$, 
then the corresponding 
$i \in {\bf Z}$,  
such that 
$\beta^{-i}$
counts the number of flat steps starting at the origin, satisfies 
$i \in 
\left[
\frac k2, N- \frac k2
\right]$.
We 
decompose 
$Tr H_{\alpha, N}^k$
as 
\beq
Tr H_{\alpha, N}^k
&=&
\sum_i
(H^k_{\alpha, N})_{ii}
=
\sum_i
\sum_{\beta}
\sharp
{\cal C}_i^k (\beta) 
V^{\beta}
\\
&=&
\left(
\sum_{
\iota(\beta) \in [k, N-k]
}
+
\sum_{
\iota(\beta) \in [k, N-k]^c
}
\right)
\sum_i
\sharp
{\cal C}_i^k (\beta) 
V^{\beta}
\\
&=:&
(1) + (2).
\eeq
Since 
${\cal P}^k (\beta)
=
\bigcup_{i \in {\bf Z}}
{\cal P}_i^k (\beta)$
is a disjoint union, 
$p^k (\beta)
=
\sum_i
\sharp 
{\cal P}_i^k (\beta)$
so that we have 
\beq
(1)
&=&
\sum_{
\iota(\beta) \in [k, N-k]
}
\sum_i
\sharp
{\cal P}_i^k (\beta) 
V^{\beta}
=
\sum_{
\iota(\beta) \in [k, N-k]
}
p^k (\beta) 
V^{\beta}. 
\eeq
\QED
\end{proof}
At the end of this subsection, 
we give some estimates on the number of the specific paths. 
\\

%
%%%%%
{\bf Lemma 2.4 }
\beq
&(1)& \quad
\sum_{
\substack{
%\beta : \iota(\beta) = 0, \\
\beta \, : \, |\beta| = j
}
}
p^l (\beta)
%{\bf E}[ X^{\beta} ]
\le 
\left(
\begin{array}{c}
l \\ j
\end{array}
\right)
\left(
\begin{array}{c}
l-j \\ \frac {l-j}{2}
\end{array}
\right)
%
%C_X^j
\\
&(2)& \quad
\sum_{j=0}^l
\sum_{
\substack{
\beta : 
%\iota(\beta) = 0, \\
|\beta| = j
}
}
p^l (\beta)
C_X^j
\le
(C_X + 2)^l
\\
&(3)& \quad
a^k (\delta)
\le
p^k (\delta)
=
\frac {1}{\sqrt{2 \pi}}
\sqrt{k}
%\frac {1}{\sqrt{k-1}}
\cdot
2^{k}
(1 + o(1)), 
\quad
k \to \infty
\eeq
%
%
%%%%%
%
\begin{proof}
(1)
\beq
\sum_{
\substack{
%\iota(\beta) = 0, \\
\beta \, : \, |\beta| = j
}
}
p^l (\beta)
&\le&
\sharp
\left\{
{\bf y}
=
(y_0, y_1, \cdots, y_l) \in {\cal P}^l
\, \middle| \,
y_0 = y_l = 0, 
\;
\sharp \{ 
\mbox{flat steps}
\} = j
\right\}
\\
& \le &
\left(
\begin{array}{c}
l \\ j
\end{array}
\right)
\left(
\begin{array}{c}
l-j \\ \frac {l-j}{2}
\end{array}
\right)
\eeq
where 
$\left(
\begin{array}{c}
l \\ j
\end{array}
\right)$
is equal to the number of ways of choosing the flat steps, 
and 
$\left(
\begin{array}{c}
l-j \\ \frac {l-j}{2}
\end{array}
\right)$
is equal to the number of path of length 
$l-j$
starting from and coming back to the origin.
(2)
We use 
(1)
and the multinomial theorem to compute
\beq
\sum_{j=0}^l
\sum_{
\substack{
%\iota(\beta) = 0, \\
\beta \, : \, |\beta| = j
}
}
p^l (\beta)
C_X^j
& \le &
\sum_{j=0}^l
\frac {
l! 
}
{
j!
\cdot
\left[
\left(
\frac {l-j}{2}
\right)!
\right]^2
}
C_X^j
\le
\sum_{
\substack{
i, j, k \\
i + j + k = l
}
}^l
\frac {
l! 
}
{
j!
\cdot
i!
\cdot
k!
}
1^i
\cdot
1^k
\cdot
C_X^j
\\
&=&
(1 + 1 + C_X)^l
=
(C_X + 2)^l.
\eeq
(3)
The first inequality follows from 
${\cal C}^k_i (\beta) 
\subset 
{\cal P}^k_i (\beta)$.
For the second one,
we use the following equations
\beq
p^k (\delta)
&=&
\frac {k!}
{
\left[
\left(
\frac {k-1}{2}
\right)!
\right]^2
},
%\cdot
%1 (k \, : \, odd), 
\quad
k : odd
\\
\left(
\begin{array}{c}
n \\ n/2
\end{array}
\right)
&=&
\frac {1}{\sqrt{2 \pi}}
\frac {2^{n+1}}{\sqrt{n}}
(1 + o(1)), 
\quad
n : even, 
\;
n \to \infty.
\eeq
%
%and estimate it by Stirling's formula.
%
%
\QED
\end{proof}
%
%
%
%%%%%%%%%%%%%%%%%%%%%%%%%%%%%%%%%%%%
\subsection{Some Lemmas for the proof of Theorem 1}
We prepare
some lemmas which are used to show the convergence of the fluctuation.\\
{\bf Lemma 2.5 }\\
{\it 
For given 
$j, \beta$
we can find 
$a^j (\beta)$
such that
\beq
&&
\sum_{j=0}^{\infty}
c_j
\sum_{
\substack{
\iota (\beta) \in [1, N] \\
|\beta| = \ell
}
}
(
a_N^j (\beta) - p^j (\beta)
)
\left(
V^{\beta} - {\bf E}[ V^{\beta} ]
\right)
\\
&&
\stackrel{N \to \infty}{\to}
\sum_{j=0}^{\infty}
c_j
\sum_{
\substack{
\iota (\beta) \in [1, j/2] \\
|\beta| = \ell
}
}
(
a^j (\beta) - p^j (\beta)
)
\left(
V^{\beta} - {\bf E}[ V^{\beta} ]
\right), 
\quad
a.s.
\eeq
}
\begin{proof}
We first
estimate the absolute value of the quantities in question, to apply the dominated convergence theorem. 
Note that 
$a_N^j (\beta) = p^j (\beta)$
if 
$\iota (\beta)
\in
[j/2, N - j/2]$ 
which yields
\beq
&&
c_j
\sum_{
\substack{
\iota (\beta) \in [1, N] \\
|\beta| = \ell
}
}
(
a_N^j (\beta) - p^j (\beta)
)
\left(
V^{\beta} - {\bf E}[ V^{\beta} ]
\right)
\\
&=&
c_j
\left(
\sum_{
\substack{
\iota (\beta) \in [1, \frac j2] \\
|\beta| = \ell
}
}
+
\sum_{
\substack{
\iota (\beta) \in [N-\frac j2, N] \\
|\beta| = \ell
}
}
\right)
(
a_N^j (\beta) - p^j (\beta)
)
\left(
V^{\beta} - {\bf E}[ V^{\beta} ]
\right).
\eeq
Then by Lemma 2.4, 
we have
\beq
&&
|c_j|
\sum_{
\substack{
\iota (\beta) \in [1, N] \\
|\beta| = \ell
}
}
|
a_N^j (\beta) - p^j (\beta)
|
\left|
V^{\beta} - {\bf E}[ V^{\beta} ]
\right|
\\
&=&
|c_j|
\left(
\sum_{i=1}^{j/2}
+
\sum_{i=N - \frac j2}^N
\right)
\sum_{
\substack{
\iota(\beta) = 0 \\
|\beta| = \ell
}
}
|
a_N^j (\beta) - p^j (\beta)
|
\left|
V^{\beta^i} - {\bf E}[ V^{\beta^i} ]
\right|
\\
& \le &
|c_j|
j
\binom{j}{\ell}
\dbinom{j - \ell}{\frac{j - \ell}{2}}
2 C_X^{\ell}
%\\
%
%& \le &
\le
2 |c_j|
j (2 + C_X)^j.
\eeq
Here we have 
$
2
\sum_{j=0}^{\infty}
|c_j|
j (2 + C_X)^j
< \infty
$
by assumption.
On the other hand, if 
$j < N$, 
$\iota(\beta) \in 
[1, j/2]$, 
$a_N^j (\beta)$
does not depend on 
$N$ : 
$a_N^j (\beta)=: a^j (\beta)$.
Moreover, 
\beq
&&
|c_j|
\sum_{
\substack{
\iota (\beta) \in [N - \frac j2, N] \\
|\beta| = \ell
}
}
|
a_N^j (\beta) - p^j (\beta)
|
\left|
V^{\beta} - {\bf E}[ V^{\beta} ]
\right|
\\
&\le&
|c_j|
\sum_{i=N - \frac j2}^N
\sum_{
\substack{
\iota (\beta) =0 \\
|\beta| = \ell
}
}
|
a_N^j (\beta) - p^j (\beta)
|
\left|
V^{\beta^i} - {\bf E}[ V^{\beta^i} ]
\right|
\\
& \le &
|c_j|
j
\binom{j}{\ell}
\dbinom{j - \ell}{\frac{j - \ell}{2}}
\frac {
2 C_X^{\ell}
}
{
\left(
N - \frac j2
\right)^{\alpha \ell}
}
\stackrel{N \to \infty}{\to} 0.
\eeq
Therefore 
the conclusion follows from the dominated convergence theorem.
\QED
\end{proof}
{\bf Lemma 2.6 }\\
{\it 
\beq
&&
\sum_{j=0}^{\infty}
c_j
\sum_{
\substack{
\iota (\beta) \in [1, N] \\
|\beta| \ge \ell_0
}
}
(
a_N^j (\beta) - p^j (\beta)
)
\left(
V^{\beta} - {\bf E}[ V^{\beta} ]
\right)
\\
&&
\stackrel{N \to \infty}{\to}
\sum_{j=0}^{\infty}
c_j
\sum_{
\substack{
\iota (\beta) \in [1, j/2] \\
|\beta| \ge \ell_0
}
}
(
a^j (\beta) - p^j (\beta)
)
\left(
V^{\beta} - {\bf E}[ V^{\beta} ]
\right), 
\quad
a.s.
\eeq
}
\begin{proof}
It suffices to make sure that the condition to apply the dominated convergence theorem is satisfied : then we sum up the result of Lemma 2.5 with respect to 
$\ell \ge \ell_0$.
The point 
is that, for fixed
$j$, 
$\sum_{\ell \ge \ell_0}$
becomes a finite sum.
In fact, 
\beq
&&
|c_j|
\sum_{
\substack{
\iota (\beta) \in [1, N] \\
|\beta| \ge \ell_0
}
}
|
a_N^j (\beta) - p^j (\beta)
|
\left|
V^{\beta} - {\bf E}[ V^{\beta} ]
\right|
\\
& = &
|c_j|
\sum_{\ell = \ell_0}^j
\sum_{
\substack{
\iota (\beta) \in [1, N] \\
|\beta| = \ell
}
}
|
a_N^j (\beta) - p^j (\beta)
|
\left|
V^{\beta} - {\bf E}[ V^{\beta} ]
\right|
\\
& \le &
|c_j|
\sum_{\ell = \ell_0}^j
j
\binom{j}{\ell}
\dbinom{j - \ell}{\frac{j - \ell}{2}}
2 C_X^{\ell}
%\\
%
%& \le &
\le
2 |c_j|
j^2 (2 + C_X)^j.
\eeq
\QED
\end{proof}
{\bf Lemma 2.7}\\
{\it 
Suppose
$\alpha \ell_0 > 1$.
Then 
\beq
\sum_{j=0}^{\infty}
c_j
\sum_{
\substack{
\iota (\beta) \in [1, N] \\
|\beta| \ge \ell_0
}
}
p^j (\beta)
\left(
V^{\beta} - {\bf E}[ V^{\beta} ]
\right)
\eeq
converges as 
$N \to \infty$ 
almost surely.}
\begin{proof}
Since 
\beq
\sum_{j=0}^{\infty}
c_j
\sum_{
\substack{
\iota (\beta) \in [1, N] \\
|\beta| \ge \ell_0
}
}
p^j (\beta)
\left(
V^{\beta} - {\bf E}[ V^{\beta} ]
\right)
&=&
\sum_{j=0}^{\infty}
\sum_{i=1}^N
c_j
\sum_{
\substack{
\iota (\beta) =0 \\
|\beta| \ge \ell_0
}
}
p^j (\beta)
\left(
V^{\beta^i} - {\bf E}[ V^{\beta^i} ]
\right), 
\eeq
it suffices to show that RHS converges absolutely.
In fact, 
\beq
\sum_{j=0}^{\infty}
\sum_{i=1}^{\infty}
|c_j|
\sum_{
\substack{
\iota (\beta) =0 \\
|\beta| \ge \ell_0
}
}
p^j (\beta)
\left|
V^{\beta^i} - {\bf E}[ V^{\beta^i} ]
\right|
%\\
%
& \le &
\sum_{j=0}^{\infty}
\sum_{i=1}^{\infty}
|c_j|
\sum_{\ell = \ell_0}^j
\sum_{
\substack{
\iota (\beta) =0 \\
|\beta| = \ell
}
}
p^j (\beta)
\frac {
2C_X^{\ell}
}
{
i^{\alpha \ell}
}
\\
& \le &
\sum_{j=0}^{\infty}
\sum_{i=1}^{\infty}
|c_j|
\sum_{\ell = \ell_0}^j
\binom{j}{\ell}
\dbinom{j - \ell}{\frac{j - \ell}{2}}
\frac {2 C_X^{\ell}}{i^{\alpha \ell_0}}
\\
& \le &
2
\sum_{j=0}^{\infty}
|c_j|
(2 + C_X)^j
\sum_{i=1}^{\infty}
\frac {1}{i^{\alpha \ell_0}}
< 
\infty.
%\quad
%a.s.
\eeq
\QED
\end{proof}
{\bf Lemma 2.8}\\
{\it 
Suppose 
$\alpha \ell > \frac 12$.
Then 
\beq
\sum_{
\substack{
\iota (\beta) \in [1, N] \\
|\beta| = \ell
}
}
\sum_{j=0}^{\infty}
c_j
p^j (\beta)
\left(
V^{\beta} - {\bf E}[ V^{\beta} ]
\right)
\eeq
converges as 
$N \to \infty$ 
in probability.
}
\begin{proof}
We show 
that the quantity in question is Cauchy in probability.
\beq
&&
{\bf E}
\left[
\left|
\sum_{
\substack{
\iota (\beta) \in [N+1, M] \\
|\beta| = \ell
}
}
\sum_{j=0}^{\infty}
c_j
p^j (\beta)
\left(
V^{\beta} - {\bf E}[ V^{\beta} ]
\right)
\right|^2
\right]
\\
&=&
{\bf E}
\left[
\left|
\sum_{i=N+1}^M
\sum_{
\substack{
\iota (\beta) =0 \\
|\beta| = \ell
}
}
\sum_{j=0}^{\infty}
c_j
p^j (\beta)
\left(
V^{\beta^i} - {\bf E}[ V^{\beta^i} ]
\right)
\right|^2
\right]
\\
&=&
{\bf E}
\Biggl[
\sum_{i_1 = N+1}^M
\sum_{i_2 = N+1}^M
\sum_{
\substack{
\iota(\beta) = 0 \\
|\beta| = \ell
}
}
\sum_{
\substack{
\iota(\gamma) = 0 \\
|\gamma| = \ell
}
}
\left(
\sum_{j \ge 0}
c_j p^j (\beta)
\right)
\left(
\sum_{j' \ge 0}
c_{j'} p^{j'} (\gamma)
\right)
\\
&& 
\hspace{10em}
\times
\left(
V^{\beta^{i_1}} - {\bf E}[ V^{\beta^{i_1}} ]
\right)
\left(
V^{\gamma^{i_2}} - {\bf E}[ V^{\gamma^{i_2}} ]
\right)
\Biggl]
\\
&=&
\sum_{i_1 = N+1}^M
\sum_{i_2 = N+1}^M
\sum_{
\substack{
\iota(\beta) = 0 \\
|\beta| = \ell
}
}
\sum_{
\substack{
\iota(\gamma) = 0 \\
|\gamma| = \ell
}
}
\left(
\sum_{j \ge 0}
c_j p^j (\beta)
\right)
\left(
\sum_{j' \ge 0}
c_{j'} p^{j'} (\gamma)
\right)
Cov
\left(
V^{\beta^{i_1}}, V^{\gamma^{i_2}}
\right).
\eeq
We 
remark that, the estimate 
\beq
p^j (\beta)
\le
\binom{j}{\ell}
\binom{j - \ell}{
\frac {j - \ell}{2}
}
\le
\binom{j}{\ell}
\cdot
2^{j - \ell}
\eeq
guarantee the absolute convergence of 
$\sum_{j \ge 0}
c_j p^j (\beta)$.
%
%Since 
%$j_1 < \ell$
%implies 
%$p^{j_1}(\beta) = 0$, 
%we have 
We further compute
\beq
&=&
\sum_{
\substack{
\iota(\beta) = 0 \\
|\beta| = \ell
}
}
\sum_{
\substack{
\iota(\gamma) = 0 \\
|\gamma| = \ell
}
}
\left(
\sum_{j_1 \ge 0}
c_{j_1} p^{j_1} (\beta)
\right)
\left(
\sum_{j_2 \ge 0}
c_{j_2} p^{j_2} (\gamma)
\right)
\sum_{i_1 = N+1}^M
\sum_{i_2 = N+1}^M
Cov
\left(
V^{\beta^{i_1}}, V^{\gamma^{i_2}}
\right)
\\
&=&
\sum_{
\substack{
\iota(\beta) = 0 \\
|\beta| = \ell
}
}
\sum_{
\substack{
\iota(\gamma) = 0 \\
|\gamma| = \ell
}
}
\left(
\sum_{j_1 \ge 0}
c_{j_1} p^{j_1} (\beta)
\right)
\left(
\sum_{j_2 \ge 0}
c_{j_2} p^{j_2} (\gamma)
\right)
\\
&&
\hspace{5em}
\times
\left(
\sum_{i_1 = N+1}^M
\sum_{
\substack{
i_2 \in [N+1, M] \\
\cap [i_1, i_1 + j_1]
}
}
+
\sum_{i_2 = N+1}^M
\sum_{
\substack{
i_1 \in [N+1, M] \\
\cap [i_2+1, i_2 + j_2]
}
}
\right)
Cov
\left(
V^{\beta^{i_1}}, V^{\gamma^{i_2}}
\right)
\\
& \le &
2 
\sum_{
\substack{
\iota(\beta) = 0 \\
|\beta| = \ell
}
}
\sum_{
\substack{
\iota(\gamma) = 0 \\
|\gamma| = \ell
}
}
\left(
\sum_{j_1 \ge 0}
|c_{j_1}| p^{j_1} (\beta)
\right)
\left(
\sum_{j_2 \ge 0}
|c_{j_2}| p^{j_2} (\gamma)
\right)
\sum_{i_1 = N+1}^M
\sum_{
\substack{
i_2 \in [N+1, M] \\
\cap [i_1, i_1 + j_1]
}
}
\frac {
\left|
Cov
\left(
X^{\beta^{i_1}}, X^{\gamma^{i_2}}
\right)
\right|
}
{
\prod_h 
h^{ \alpha (\beta_h^{i_1} + \gamma_h^{i_2})}
}.
\eeq
By using  
\beq
\prod_h 
h^{ \alpha (\beta_h^{i_1} + \gamma_h^{i_2})}
& \ge &
i_1^{ \alpha(|\beta| + | \gamma |) }
=
i_1^{2 \alpha \ell}, 
\qquad
%\\
%
\left|
Cov
\left(
X^{\beta^{i_1}}, X^{\gamma^{i_2}}
\right)
\right|
\le 
(2 C_X)^{2 \ell}
\eeq
we have 
\beq
& \le &
2 
\sum_{
\substack{
\iota(\beta) = 0 \\
|\beta| = \ell
}
}
\sum_{
\substack{
\iota(\gamma) = 0 \\
|\gamma| = \ell
}
}
\sum_{j_1 \ge 0}
|c_{j_1}| p^{j_1} (\beta)
\sum_{j_2 \ge 0}
|c_{j_2}| p^{j_2} (\gamma)
\sum_{i_1 = N+1}^M
\sum_{
\substack{
i_2 \in [N+1, M] \\
\cap [i_1, i_1 + j_1]
}
}
\frac { 
4 C_X^{2 \ell}
}
{
i_1^{2 \alpha \ell}
}
\\
& \le &
8
\left(
\sum_{j_1 \ge 0}
|c_{j_1}|
(j_1 + 1) 
\sum_{
\substack{
\iota(\beta) = 0 \\
|\beta| = \ell
}
}
p^{j_1} (\beta)
\right)
\left(
\sum_{j_2 \ge 0}
|c_{j_2}|
\sum_{
\substack{
\iota(\gamma) = 0 \\
|\gamma| = \ell
}
} 
p^{j_2} (\gamma)
\right)
C_X^{2\ell}
\sum_{i_1 = N+1}^M
\frac {1}
{
i_1^{2 \alpha \ell}
}.
\eeq
Here we notice that 
\beq
\sum_{j_1 \ge 0}
|c_{j_1}|
(j_1 + 1) 
\sum_{
\substack{
\iota(\beta) = 0 \\
|\beta| = \ell
}
}
p^{j_1} (\beta)
\le
\sum_{j_1 \ge 0}
|c_{j_1}|
(j_1 + 1) 
\binom{j}{\ell} 2^{j_1 - \ell}
< \infty.
\eeq
Thus 
by Chebyshev's inequality, 
\beq
&&
{\bf P}
\left(
\left|
\sum_{j=0}^{\infty}
c_j
\sum_{
\substack{
\iota (\beta) \in [1, N] \\
|\beta| = \ell
}
}
p^j (\beta)
\left(
V^{\beta} - {\bf E}[ V^{\beta} ]
\right)
-
\sum_{j=0}^{\infty}
c_j
\sum_{
\substack{
\iota (\beta) \in [1, M] \\
|\beta| = \ell
}
}
p^j (\beta)
\left(
V^{\beta} - {\bf E}[ V^{\beta} ]
\right)
\right|
\ge
\epsilon
\right)
\\
& \le &
\epsilon^{-2}
{\bf E}
\left[
\left|
\sum_{j=0}^{\infty}
c_j
\sum_{
\substack{
\iota (\beta) \in [1, N] \\
|\beta| = \ell
}
}
p^j (\beta)
\left(
V^{\beta} - {\bf E}[ V^{\beta} ]
\right)
-
\sum_{j=0}^{\infty}
c_j
\sum_{
\substack{
\iota (\beta) \in [1, M] \\
|\beta| = \ell
}
}
p^j (\beta)
\left(
V^{\beta} - {\bf E}[ V^{\beta} ]
\right)
\right|^2
\right]
\\
& \to &
0, 
\quad
N, M \to \infty
\eeq
\QED
\end{proof}
Before
ending this section, we show below that the ``limit" of the random variable in Lemma 2.8 really converges as a random series.
This fact 
is not used in this paper, but we include here for completeness.\\

{\bf Lemma 2.9}
{\it 
Suppose
$\alpha \ell_0 > \frac 12$.
Then 
the random series 
\beq
\sum_{j=0}^{\infty}
c_j
\sum_{
\substack{
\iota(\beta) \in [1, \infty) \\
|\beta| \ge \ell_0
}
}
p^j (\beta)
\left(
V^{\beta} - {\bf E}[ V^{\beta} ] 
\right)
\eeq
converges and define a random variable.
}\\
{\bf Remark }
Lemma 2.9
says that, the random series 
\beq
\sum_{j=0}^{\infty}
c_j
\sum_{i=1}^{\infty}
\sum_{
\substack{
\iota(\beta)=0 \\
|\beta| \ge \ell_0
}
}
p^j (\beta)
\left(
V^{\beta^i} - {\bf E}[ V^{\beta^i} ] 
\right)
\eeq
converges almost surely, if one takes 
$\sum_{i=1}^{\infty}$
first and then 
$\sum_{j=0}^{\infty}$.
Since 
we are not able to show the absolute convergence, we do not know whether 
a.s. convergence holds if we change the order of summation.
\begin{proof}
We first
note that, by Lemma 4.2 in \cite{B}, 
The random series 
\beq
\sum_{
\substack{
\iota(\beta) \in [1, N] \\
|\beta| \ge \ell_0
}
}
p^j (\beta) 
\left(
V^{\beta} - {\bf E}[ V^{\beta} ]
\right)
=
\sum_{i=1}^N
\sum_{
\substack{
\iota(\beta)=0 \\
|\beta| \ge \ell_0
}
}
p^j (\beta) 
\left(
V^{\beta^i} - {\bf E}[ V^{\beta^i} ]
\right)
%\quad\cdots (*)
\eeq
converges a.s. and define a random variable.
\beq
\sum_{
\substack{
\iota(\beta) \in [1, \infty) \\
|\beta| \ge \ell_0
}
}
p^j (\beta)
\left(
V^{\beta} - {\bf E}[ V^{\beta} ] 
\right)
=
\sum_{i=1}^{\infty}
\sum_{
\substack{
\iota(\beta)=0 \\
|\beta| \ge \ell_0
}
}
p^j (\beta) 
\left(
V^{\beta^i} - {\bf E}[ V^{\beta^i} ]
\right)
\eeq
with finite variance of which we estimate below. 
\beq
&&
Var
\left(
\sum_{i=1}^{\infty}
\sum_{
\substack{
\iota(\beta) = 0 \\
|\beta| \ge \ell_0
}
}
p^j (\beta) 
\left(
V^{\beta^i} - {\bf E}[ V^{\beta^i} ]
\right)
\right)
\\
& \le &
\liminf_{N \to \infty}
{\bf E}
\left[
\left(
\sum_{i=1}^{N}
\sum_{
\substack{
\iota(\beta) = 0 \\
|\beta| \ge \ell_0
}
}
p^j (\beta)
\left(
V^{\beta^i} - {\bf E}[ V^{\beta^i} ]
\right)
\right)^2
\right]
\\
& \le &
\sum_{
\substack{
\iota(\beta) = 0 \\
|\beta| \ge \ell_0
}
}
p^j (\beta)
\sum_{
\substack{
\iota(\gamma) = 0 \\
|\gamma| \ge \ell_0
}
}
p^j (\gamma)
\limsup_{N \to \infty}
\sum_{i=1}^{N}
\sum_{i'=1}^{N}
\left|
Cov
\left(
\left(
V^{\beta^i} - {\bf E}[ V^{\beta^i} ]
\right), 
\left(
V^{\gamma^{i'}} - {\bf E}[ V^{\gamma^{i'}} ]
\right)
\right)
\right|
\eeq
Since 
the covariance vanishes if 
$|i - i'| > \frac j2$, 
\beq
\sum_{i=1}^{N}
\sum_{i'=1}^{N}
\left|
Cov
\left(
\left(
V^{\beta^i} - {\bf E}[ V^{\beta^i} ]
\right), 
\left(
V^{\gamma^{i'}} - {\bf E}[ V^{\gamma^{i'}} ]
\right)
\right)
\right|
\le
\sum_{i=1}^N
j
\cdot
\frac {
4 C_X^{ |\beta| + |\gamma| } 
}
{
i^{2\alpha \ell_0}
}
\eeq
so that by Lemma 2.4, 
\beq
&&
\sum_{
\substack{
\iota(\beta) = 0 \\
|\beta| \ge \ell_0
}
}
p^j (\beta)
\sum_{
\substack{
\iota(\gamma) = 0 \\
|\gamma| \ge \ell_0
}
}
p^j (\gamma)
\limsup_{N \to \infty}
\sum_{i=1}^{N}
\sum_{i'=1}^{N}
\left|
Cov
\left(
\left(
V^{\beta^i} - {\bf E}[ V^{\beta^i} ]
\right), 
\left(
V^{\gamma^{i'}} - {\bf E}[ V^{\gamma^{i'}} ]
\right)
\right)
\right|
\\
& \le &
4
\sum_{
\substack{
\iota(\beta) = 0 \\
|\beta| \ge \ell_0
}
}
p^j (\beta)
\sum_{
\substack{
\iota(\gamma) = 0 \\
|\gamma| \ge \ell_0
}
}
p^j (\gamma)
j
C_X^{ |\beta| + |\gamma| }
\sum_{i=1}^{\infty}
\frac {
1
} 
{
i^{2\alpha \ell_0}
}
%\\
%
%& \le &
\le
4 
j
(C_X+2)^{2j} 
\sum_{i=1}^{\infty}
\frac {
1
} 
{
i^{2\alpha \ell_0}
}.
\eeq
By Chebyshev's inequality, 
\beq
{\bf P}
\left(
\left|
\sum_{i=1}^{\infty}
\sum_{
\substack{
\iota(\beta) = 0 \\
|\beta| \ge \ell_0
}
}
p^j (\beta) 
\left(
V^{\beta^i} - {\bf E}[ V^{\beta^i} ]
\right)
\right|
>
j^{\delta}
(C_X + 2)^j
\right)
\le
4 
j^{1 - 2 \delta}
%(C_X+2)^{2j} 
\sum_{i=1}^{\infty}
\frac {
1
} 
{
i^{2\alpha \ell_0}
}
\eeq
which is summable with respect to 
$j$
if 
$\delta > 1$.
Thus 
by Borel-Cantelli lemma, 
\beq
\left|
\sum_{i=1}^{\infty}
\sum_{
\substack{
\iota(\beta) = 0 \\
|\beta| \ge \ell_0
}
}
p^j (\beta) 
\left(
V^{\beta^i} - {\bf E}[ V^{\beta^i} ]
\right)
\right|
\le
j^{\delta}
(C_X + 2)^j
\eeq
for sufficiently large 
$j$, 
almost surely. 
Therefore 
\beq
\sum_{j=0}^{\infty}
c_j
\sum_{i=1}^{\infty}
\sum_{
\substack{
\iota(\beta) = 0 \\
|\beta| \ge \ell_0
}
}
p^j (\beta) 
\left(
V^{\beta^i} - {\bf E}[ V^{\beta^i} ]
\right)
\eeq
converges almost surely. 
\QED
\end{proof}
%

%
%%%%%%%%%%%%%%%%%%%%%%%%%%%%%%%%%%%%%%%%%%%%%%%%%%%%%%%%%%%%
\section{Proof of Theorem 1}
We prove 
A(1), A(2),
$\cdots$, 
C(2) 
in Theorem 1 respectively in each 
subsection below.
%
%
%
%%%%%%%%%%%%%%%%%%%%%%%%%%%%%%%%%%%%
\subsection{Proof of A(1)}
Since 
$\sigma (H_{\alpha, N})
\subset
[-C_X-2, C_X+2]$, 
and since 
$r(f) > C_X + 2$, 
the power series 
$f(H)
=
\sum_{l \ge 0}
a_l H^l$
converges in the operator norm topology.
Hence 
$Tr f(H)
=
\sum_{l \ge 0}
c_l 
Tr (H^l)$
is absolutely convergent so that by Fubini theorem, 
\beq
{\bf E}[
Tr f(H)
]
&=&
\sum_{l \ge 0}
c_l
{\bf E}[
Tr (H^l)]. 
\eeq
Letting 
$a_{\beta}^j
:=
[V^{\beta}]
\left(
Tr (H_{\alpha, N}^j)
\right)$, 
we compute 
\beq
Tr f(H_{\alpha, N})
-
{\bf E}
\left[
Tr f(H_{\alpha, N})
\right]
&=&
\sum_{j=0}^{\infty}
c_j
\left\{
Tr 
\left(
H_{\alpha, N}^j
\right)
-
{\bf E}
\left[
Tr 
\left(
H_{\alpha, N}^j
\right)
\right]
\right\}
\\
&=&
\sum_{j=0}^{\infty}
c_j
\left(
\sum_{ | \beta | = 1}
+
\sum_{ | \beta | \ge 2}
\right)
a_{\beta}^j
\left(
V_{\alpha}^{\beta}
-
{\bf E}
\left[
V_{\alpha}^{\beta}
\right]
\right)
\\
& =: &
I + II. 
\eeq
Since 
$|\beta| = 1$
implies 
$\beta = \delta^n$, 
for some 
$n = 1, 2, \cdots, N$, 
we have 
$I = \sum_{n=1}^N
\sum_{j=0}^{\infty}
c_j
a_{ \delta^n }^j
\dfrac {X_n}{n^{\alpha}}$.
Here we use \\
%

%%%%%
{\bf Lemma 3.1
(\cite{D} Theorem 2.5.6)}\\
{\it 
Suppose 
$\{ Y_n \}_n$
are independent with 
${\bf E}[ Y_n ] = 0$.
If 
$\sum_n Var (Y_n) < \infty$, 
then 
$\sum_n Y_n$
is convergent a.s.
}
\\

Put
$Y_n := 
\sum_{j=0}^{\infty}
c_j
a_{ \delta^n }^j
\dfrac {X_n}{n^{\alpha}}$.
Then by 
Lemma 2.4(3), 
\beq
Var (Y_n)
& \le &
\left(
\sum_{j=0}^{\infty}
c_j
a_{\delta^n}^j
\right)^2
\frac { \eta^2 }{n^{2 \alpha}}
\le
(Const.)
\left(
\sum_{j=0}^{\infty}
c_j
%a_{\delta^i}^j
(2 + C_X)^j
%3^j
\right)^2
\frac { \eta^2 }{n^{2 \alpha}}
\eeq
so that 
$\sum_n Var (Y_n) < \infty$
implying 
$I = \sum_n Y_n$
is convergent a.s.
\\
For 
$II$, 
the relevant paths have more than one flat steps which produces the convergent factor. 
In fact, 
by Lemma 2.4(2), 
\beq
| II |
&\le&
\sum_{j=0}^{\infty}
|c_j|
%\sum_{ | \beta | \ge 2}
\sum_{i=1}^N
\sum_{ | \beta | \ge 2 
%\iota(\beta) = 0 
}
a_{\beta^i}^j
\left|
\frac {
X^{\beta} - {\bf E}[ X^{\beta} ]
}
{
\prod_{n \in {\bf N}}
n^{\alpha \beta_n}
}
\right|
\\
& \le &
\sum_{j=0}^{\infty}
|c_j|
%\sum_{ | \beta | \ge 2}
\sum_{i=1}^N
\sum_{ | \beta | \ge 2 
%\iota(\beta) = 0 
}
%a_{\beta^i}^j
p^j (\beta)
\frac {
2 (C_X)^{|\beta|}
%X^{\beta} - {\bf E}[ X^{\beta} ]
}
{
%\prod_{n \in {\bf N}}
%n^{\alpha \beta_n}
i^{2 \alpha}
}
\\
& \le &
(Const.)
\sum_{j=0}^{\infty}
|c_j|
(C_X + 2)^j
%\sum_{ | \beta | \ge 2}
\sum_{i=1}^N
%\sum_{ | \beta | \ge 2, \iota(\beta) = 0 }
%a_{\beta^i}^j
%p^j (\beta)
\frac {
1
%2 (C')^j
%X^{\beta} - {\bf E}[ X^{\beta} ]
}
{
%\prod_{n \in {\bf N}}
%n^{\alpha \beta_n}
i^{2 \alpha}
}
< 
\infty
\eeq
Therefore 
$II$ 
converges a.s.
%

%
%%%%%%%%%%%%%%%%%%%%%%%%%%%%%%%%%%%%
\subsection{Proof of A(2)}
The key lemma is : \\

%%%%%
{\bf Lemma 3.2
(\cite{Bil} Theorem 25.5) }
{\it 
Suppose that 

$\{ Y_N \}_N$, 
$\{ X_{N, k} \}$, 
$N, k \in {\bf N}$
are random variables satisfying \\
(i)
$X_{N, k} 
\stackrel{d}{\to}
X_k$, 
$N \to \infty$, 
\mbox{ for any fixed }
k
\\
(ii)
$X_k 
\stackrel{d}{\to} 
X$, 
$k \to \infty$
\\
(iii)
For any 
$\epsilon > 0$, 
%\beq
$
\lim_{ k \to \infty}
\limsup_{ N \to \infty}
{\bf P}
\left(
| Y_N - X_{N, k} | > \epsilon
\right)
= 0.
$
%\eeq
%
\\
Then 
$Y_N 
\stackrel{d}{\to}
X$.
}
\\
%
%%%%%

We set 
\beq
X_{N, k}
&:=&
\frac {
Tr f_k(H_{\alpha, N}) - 
{\bf E}[ 
Tr f_k(H_{\alpha, N})
]
}
{
g_{2 \alpha} (N)
}, 
\quad
X_k 
:=
N(0, 
\sigma_A (P_k)^2)
\\
Y_N
&:=&
\frac {
Tr f(H_{\alpha, N}) - 
{\bf E}[ 
Tr f(H_{\alpha, N})
]
}
{
g_{2 \alpha} (N)
}, 
\quad
X 
:=
N(0, 
\sigma_A (f)^2).
\eeq
to apply Lemma 3.2, where 
$f_k(x) := \sum_{j=0}^k c_j x^j$
is the polynomial by truncating the Taylor expansion of 
$f$.
To finish the proof, 
it suffices 
to check the three conditions in Lemma 3.2\\
(i)
By Theorem 1.1 in 
\cite{B}, 
we have 
$X_{N, k} \stackrel{d}{\to} X$.
\\
(ii)
$X_k 
\stackrel{d}{\to}
X$
is obvious. 
In fact,  
the characteristic function of 
$X_k$ 
converges to that of 
$X$ : 
$\exp 
\left(
- \frac 12 \sigma_A (f_k)^2 t^2
\right)
\stackrel{k \to \infty}{\to}
\exp 
\left(
- \frac 12 \sigma_A (f)^2 t^2
\right)$.\\
(iii)
As in the proof for Theorem 1(1), we can show
\begin{equation}
Var 
\left(
Y_N - X_{N, k}
\right)
=
{\bf E}
\left[
\left(
\sum_{j \ge k+1}
c_j
\frac {
\sum_{ |\beta| \ge 1}
a_{\beta}^j 
( V_{\alpha}^{\beta} - {\bf E}[ V_{\alpha}^{\beta} ] )
}
{
g_{2 \alpha } (N)
} 
\right)^2
\right]
\label{var}
%\quad\cdots (*)
\end{equation}
We further compute 
\begin{equation}
{\bf E}
\left[
\left\{
\sum_{j \ge k+1}
c_j
\sum_{ |\beta| \ge 1}
a_{\beta}^j 
( V_{\alpha}^{\beta} - {\bf E}[ V_{\alpha}^{\beta} ] )
\right\}^2
\right]
=
\sum_{j_1 \ge k+1}
\sum_{j_2 \ge k+1}
c_{j_1}
c_{j_2}
\sum_{ |\beta| \ge 1}
\sum_{ |\gamma| \ge 1}
a_{\beta}^{j_1} 
a_{\gamma}^{j_2} 
\,
\cdot
\,
Cov (V_{\alpha}^{\beta}, V_{\alpha}^{\gamma})
\label{num}
%\quad\cdots (\sharp)
\end{equation}
and moreover
\beq
&&
\sum_{ |\beta|, | \gamma | \ge 1}
a_{\beta}^{j_1} 
a_{\gamma}^{j_2} 
\,
\cdot
\,
| 
Cov (V_{\alpha}^{\beta}, V_{\alpha}^{\gamma})
|
\\
& \le &
\sum_{i=1}^N
\sum_{i'=1}^N
\sum_{ \iota (\beta) = 0}
\sum_{ \iota (\gamma) = 0}
p^{j_1}(\beta)
p^{j_2}(\gamma)
| 
Cov (V_{\alpha}^{\beta^i}, V_{\alpha}^{\gamma^{i'}})
|
\\
&=&
\sum_{ \iota (\beta) = 0}
\sum_{ \iota (\gamma) = 0}
p^{j_1}(\beta)
p^{j_2}(\gamma)
\sum_{\ell = - \infty}^{\infty}
| 
Cov (X^{\beta}, X^{\gamma^{\ell}})
|
\sum_{i \in I_{N, \ell}}
\prod_{n \in {\bf N}}
\frac {1}
{
n^{\alpha (\beta_{n-i} + \gamma_{n - i-\ell}) }
}
\eeq
where we set 
$I_{N, \ell}
:=
\left\{
i 
\, \middle| \,
1 \le i, i + \ell \le N
\right\}$.
The covariance and the sum of the products are bounded by 
\beq
&&
| Cov (X^{\beta}, X^{\gamma^{\ell}}) |
\le
\sqrt{
V (X^{\beta}) \cdot V(X^{\gamma})
}
\le
C_X^{|\beta| + |\gamma|}
\le
C_X^{j_1 + j_2}
\\
&&
\sum_{i \in I_{N, \ell}}
\prod_{n \in {\bf N}}
\frac {1}
{
n^{\alpha (\beta_{n-i} + \gamma_{n - i-\ell}) }
}
\le
\sum_{i \in I_{N, \ell}}
\frac {1}
{
i^{\alpha |\beta|}
\dot
(i + \ell)^{\alpha | \gamma |}
}
\le
\sum_{i=1}^N
\frac {1}{i^{2 \alpha}}.
\eeq
On the other hand, since 
$supp \;\beta 
\subset
\left[
0, \frac {j_1}{2}
\right]$, 
$
supp \; \gamma 
\subset
\left[
0, \frac {j_2}{2}
\right]$, 
the number of 
$\ell$'s 
such that 
$Cov (X^{\beta}, X^{\gamma^{\ell}}) \ne 0$
is at most 
$(j_1+j_2) (\le j_1 \cdot j_2)$.
Using all these estimates and Lemma 2.4 yields
\beq
&&
\sum_{ |\beta|, | \gamma | \ge 1}
\alpha_{\beta}^{j_1} 
\alpha_{\gamma}^{j_2} 
\,
\cdot
\,
| 
Cov (V_{\alpha}^{\beta}, V_{\alpha}^{\gamma})
|
\\
& \le &
\sum_{ \iota (\beta) = 0}
\sum_{ \iota (\gamma) = 0}
p^{j_1}(\beta)
p^{j_2}(\gamma)
j_1 
\cdot
j_2
\cdot
C_X^{|\beta| + |\gamma|}
%C_X^{j_1+j_2}
\sum_{i=1}^N 
\frac {1}{i^{2 \alpha}}
\\
& \le &
j_1(C_X + 2)^{j_1}
\cdot
j_2(C_X + 2)^{j_2}
\sum_{i=1}^N 
\frac {1}{i^{2 \alpha}}.
\eeq
Plugging 
this one to 
(\ref{num}) 
and then 
(\ref{var}), 
we have 
\beq
{\bf E}
\left[
\left\{
\sum_{j \ge k+1}
c_j
\sum_{ |\beta| \ge 1}
\alpha_{\beta}^j 
( V_{\alpha}^{\beta} - {\bf E}[ V_{\alpha}^{\beta} ] )
\right\}^2
\right]
& \le &
\left(
\sum_{j=k+1}^{\infty}
|c_j|
j(C_X + 2)^{j}
\right)^2
\sum_{i=1}^N 
\frac {1}{i^{2 \alpha}}, 
\\
Var 
\left(
Y_N - X_{N, k}
\right)
& \le &
D_k^2
\frac {
\sum_{i=1}^N 
\frac {1}{i^{2 \alpha}}
}
{
g_{2 \alpha} (N)
}.
\eeq
where we set 
$D_k:=
\sum_{j=k+1}^{\infty}
|c_j|
j(C_X + 2)^{j}$.
Therefore, 
\beq
\lim_{ k \to \infty}
\limsup_{ N \to \infty}
P (
|X_{N,k} - Y_N | \ge \epsilon
)
\le
\lim_{ k \to \infty}
\limsup_{ N \to \infty}
D_k^2
\frac {
\sum_{i=1}^N 
\frac {1}{i^{2 \alpha}}
}
{
g_{2 \alpha} (N)
}
= 0.
%\le
%\frac {M}{\epsilon^2} 
%\stackrel{k \to \infty}{\to} 0
\eeq
\QED
%
%%%%%%%%%%%%%
%
%%%%%%%%%%%%%%%%%%%%%%%%%%%%%%%%%%%%
\subsection{Proof of B(1)}
By assumption
$c_j = 0$
for 
$j$ : odd
so that if 
$| \beta | \notin 2 {\bf N}$, 
${\cal C}_{i, j} (\beta) = \emptyset$, 
$\forall i$.
We then compute 
\beq
&&
\sum_{j=0}^{\infty}
c_j
\left\{
Tr (H_{\alpha, N}^j)
-
{\bf E}[
Tr (H_{\alpha, N}^j)
]
\right\}
\\
&=&
\sum_{j=0}^{\infty}
c_j
\sum_{
\substack{
\iota(\beta) \in [1, N] \\
|\beta| = 2
}
}
p^j (\beta)
\left(
V^{\beta} - {\bf E}[ V^{\beta} ] 
\right)
+
\sum_{j=0}^{\infty}
c_j
\sum_{
\substack{
\iota(\beta) \in [1, N] \\
|\beta| \ge 4
}
}
p^j (\beta)
\left(
V^{\beta} - {\bf E}[ V^{\beta} ] 
\right)
\\
&&
+
\sum_{j=0}^{\infty}
c_j
\sum_{
\substack{
\iota(\beta) \in [1, N] \\
|\beta| \ge 2
}
}
\left(
a_{\beta}^{j, N} - p^j (\beta) 
\right)
\left(
V^{\beta} - {\bf E}[ V^{\beta} ] 
\right)
\\
&=:&
I + II + III.
\eeq
Here 
we apply 
Lemma 2.8 to 
$I$, 
Lemma 2.7 to 
$II$, 
and Lemma 2.6 
to 
$III$, 
to have the desired result.
\QED
%
%
%%%%%%%%%%%%%%%%%%%%%%%%%%%%%%%%%%%%
\subsection{Proof of B(2)}
Proof of 
B(2) 
and
C(2)
is similar to that of 
A(1) : 
let 
\beq
X_{N, k}
&:=&
\frac {
Tr f_k (H_{\alpha, N}) - 
{\bf E}[ Tr f_k (H_{\alpha, N}) ] 
}
{
g_{4 \alpha}(N)
}, 
\quad
X_k
:=
N(0, \sigma_{B} (f_k)^2 )
\\
Y_N
&:=&
\frac {
Tr f (H_{\alpha, N}) - 
{\bf E}[ Tr f (H_{\alpha, N}) ] 
}
{
g_{4 \alpha}(N)
}, 
\quad
X
:=
N(0, \sigma_{B} (f)^2 )
\eeq
and show that the conditions to apply Lemma 3.2 are satisfied.
%
%The first one 

(i)
$X_{N, k} 
\stackrel{d}{\to}
X_k$, 
$N \to \infty$
has been proved in 
\cite{B}.

(ii)
To show 
$X_k 
\stackrel{d}{\to}
X$,
let 
$\varphi_k (t)$
be the characteristic function of 
$X_k$.
Then 
\beq
\varphi_k (t)
&=&
\exp
\left(
- \frac 12
\sigma_{B} (f_k)^2 t^2
\right)
\\
&=&
\exp
\left[
- \frac 12
\left\{
\left(
\sum_{j=2}^k 
c_j p^j (2 \delta) 
\right)^2
\left(
{\bf E}[ X_1^4 ] - \eta^4
\right)
+
\sum_{s=1}^{\infty}
\left(
\sum_{j=2}^k
c_j
p^j (\delta + \delta^s)
\right)^2
\eta^4
\right\}
t^2
\right].
\eeq
To estimate 
that the second term in the exponential factor in RHS, we use 
\beq
p^j (2\delta)
&=&
j 2^{j-3}, 
\\
p^j (\delta + \delta^s)
&\le&
\frac {j (j-1)}{2}
\dbinom{j-2}{(j-2)/2}
\le
\frac {j^2}{2}
\frac {1}{\sqrt{j-2}}
2^{j-1} (1 + o(1))
\le
(Const.)
j^{3/2} 2^{j-1}, 
\eeq
and thus
\beq
\sum_{s=1}^{\infty}
\left(
\sum_{j=2}^k
c_j
p^j (\delta + \delta^s)
\right)^2
&=&
\sum_{s=1}^{\infty}
\sum_{j=2}^k
\sum_{j'=2}^k
c_j 
\cdot
c_{j'}
p^j (\delta + \delta^s)
p^{j'} (\delta + \delta^s)
\\
&=&
\sum_{j=2}^k
\sum_{j'=2}^k
c_j \cdot c_{j'}
\left(
\sum_{s=1}^{j \vee j'}
p^j (\delta + \delta^s)
p^{j'} (\delta + \delta^s)
\right)
\\
&\le&
\left(
\sum_{j=2}^k
c_j j^{5/2} 2^{j-1}
\right)^2.
\eeq
Therefore 
the RHS converges as 
$k \to \infty$
so that 
\beq
\varphi_k (t)
\stackrel{k \to \infty}{\to}
\varphi (t)
=
\exp
\left[
- \frac 12
\sigma_{B} (f)^2 t^2
\right].
\eeq
(iii)
We note that if 
$|\beta| \notin 2 {\bf N}$
${\cal C}_{i, j} (\beta) = \emptyset$, 
$\forall i$.
In particular 
$a_N^j (\beta) = 0$
for 
$|\beta| = 1$.
Hence
\beq
&&
Var
\left(
\sum_{j \ge k+1}
c_j
\frac {
Tr (H^j_{\alpha, N})
-
{\bf E}[ Tr (H^j_{\alpha, N}) ] 
}
{
g_{4 \alpha} (N)
}
\right)
\\
&=&
{\bf E}
\left[
\left\{
\sum_{j \ge k+1}
c_j
\sum_{|\beta| \ge 2}
\frac {
a_N^j (\beta)
\left(
V^{\beta} - {\bf E}[ V^{\beta} ]
\right)}
{
g_{4 \alpha} (N)
}
\right\}^2
\right]
\\
&=&
\frac {1}{ g_{4 \alpha}(N)^2 }
\sum_{j_1 \ge k+1}
\sum_{j_2 \ge k+1}
c_{j_1} \cdot c_{j_2}
\sum_{|\beta| \ge 2}
\sum_{|\gamma| \ge 2}
a_N^{j_1}(\beta)
a_N^{j_2}(\gamma)
Cov (V^{\beta}, V^{\gamma}).
\eeq
Then 
its absolute value is bounded from above by 
\beq
&&
\frac {1}{g_{4 \alpha} (N)^2 }
\sum_{j_1 \ge k+1}
\sum_{j_2 \ge k+1}
|c_{j_1}| \cdot |c_{j_2}|
\sum_{|\beta| \ge 2}
\sum_{|\gamma| \ge 2}
a_N^{j_1}(\beta)
a_N^{j_2}(\gamma)
| Cov (V^{\beta}, V^{\gamma}) |
\\
& \le &
\frac {1}{g_{4 \alpha} (N)^2 }
\sum_{j_1 \ge k+1}
\sum_{j_2 \ge k+1}
|c_{j_1}| \cdot |c_{j_2}|
\sum_{
\substack{\iota(\beta)=0 \\|\beta| \ge 2}
}
\sum_{
\substack{\iota(\gamma)=0, \\ |\gamma| \ge 2}
}
p^{j_1}(\beta)
p^{j_2}(\gamma)
\\
&& \qquad
\times
\sum_{\ell \in {\bf Z}}
| Cov (X^{\beta}, X^{\gamma^{\ell}}) |
\sum_{i \in I_{N, \ell}}
\prod_{n \in {\bf N}}
\frac {1}
{
n^{\alpha (\beta_{n-i} + \gamma_{n - i - \ell})}
}
\\
& \le &
\frac {2}{g_{4 \alpha} (N)^2 }
\sum_{j_1 \ge k+1}
\sum_{j_2 \ge k+1}
|c_{j_1}| \cdot |c_{j_2}|
\left(
j_1
\sum_{
\substack{\iota(\beta)=0 \\|\beta| \ge 2}
}
p^{j_1} (\beta) 
C_X^{|\beta|}
\right)
\left(
j_2
\sum_{
\substack{\iota(\gamma)=0 \\|\gamma| \ge 2}
}
p^{j_2} (\gamma) 
C_X^{|\gamma|}
\right)
\sum_{i=1}^N
\frac {1}{ i^{4 \alpha} }
\\
& \le &
\frac {2}{g_{4 \alpha} (N)^2 }
\left(
\sum_{j \ge k+1}
|c_j|
j
(C_X+2)^j
\right)^2
\sum_{i=1}^N
\frac {1}{ i^{4 \alpha} }.
\eeq
Take 
$\epsilon > 0$
arbitrary small.
Then 
by Chebyshev's inequality,
\beq
&&
{\bf P}
\left(
\left|
\sum_{j \ge k+1}
c_j
\frac {
Tr (H^j_{\alpha, N})
-
{\bf E}[ Tr (H^j_{\alpha, N}) ] 
}
{
g_{4 \alpha} (N)
}
\right|
\ge
\epsilon
\right)
\\
& \le &
\frac {2}{g_{4 \alpha} (N)^2 }
\left(
\sum_{j \ge k+1}
|c_j|
j
(C_X+2)^j
\right)^2
\sum_{i=1}^N
\frac {1}{ i^{4 \alpha} }
\frac {1}{\epsilon^2}.
\eeq
Therefore 
by the definition of 
$g_{4 \alpha}(N)$, 
we have 
\beq
&&
\limsup_N
{\bf P}
\left(
\left|
\sum_{j \ge k+1}
c_j
\frac {
Tr (H^j_{\alpha, N})
-
{\bf E}[ Tr (H^j_{\alpha, N}) ] 
}
{
g_{4 \alpha} (N)
}
\right|
\ge
\epsilon
\right)
\le
\frac {2}{\epsilon^2}
\left(
\sum_{j \ge k+1}
|c_j|
j
(C_X+2)^j
\right)^2
\\
&&
\lim_k
\limsup_N
{\bf P}
\left(
\left|
\sum_{j \ge k+1}
c_j
\frac {
Tr (H^j_{\alpha, N})
-
{\bf E}[ Tr (H^j_{\alpha, N}) ] 
}
{
g_{4 \alpha} (N)
}
\right|
\ge
\epsilon
\right)
=0.
\eeq
\QED
%
%
%%%%%%%%%%%%%%%%%%%%%%%%%%%%%%%%%%%%
\subsection{Proof of C(1)}
The idea 
is the same as that of the proof of 
B(1).
In fact, 
\beq
&&
Tr f(H_{\alpha, N}) - {\bf E}[ Tr f(H_{\alpha, N}) ]
\\
&=&
\sum_{j=0}^{\infty}
c_j
\Biggl\{
\Bigl(
Tr (H_{\alpha, N}^j)
-
{\bf E}
\left[
Tr (H_{\alpha, N}^j)
\right]
\Bigr)
-
\Bigl(
p^j (\delta) 
Tr (H_{\alpha, N})
-
{\bf E}
\left[
p^j (\delta) 
Tr (H_{\alpha, N})
\right]
\Bigr)
\Biggr\}
\\
&=&
\sum_{j=0}^{\infty}
\sum_{
\substack{
\iota(\beta) \in [1, N] \\
|\beta| = 3
}
}
p^j (\beta)
\left(
V^{\beta} - {\bf E}[ V^{\beta} ] 
\right)
+
\sum_{j=0}^{\infty}
c_j
\sum_{
\substack{
\iota(\beta) \in [1, N] \\
|\beta| = 5
}
}
p^j (\beta)
\left(
V^{\beta} - {\bf E}[ V^{\beta} ] 
\right)
\\
&&
+
\sum_{j=0}^{\infty}
c_j
\sum_{
\substack{
\iota(\beta) \in [1, N] \\
|\beta| \ge 7
}
}
p^j (\beta)
\left(
V^{\beta} - {\bf E}[ V^{\beta} ] 
\right)
+
\sum_{j=0}^{\infty}
c_j
\sum_{
\substack{
\iota(\beta) \in [1, N] \\
|\beta| \ge 1
}
}
\left(
a_{\beta}^{j, N} - p^j (\beta)
\right)
\left(
V^{\beta} - {\bf E}[ V^{\beta} ] 
\right)
\\
&=:&
I + II + III + IV.
\eeq
We apply 
Lemma 2.8 to 
$I, II$, 
Lemma 2.7 to 
$III$, 
and Lemma 2.6 to 
$IV$
to obtain the desired conclusion.
\QED
%
%
%%%%%%%%%%%%%%%%%%%%%%%%%%%%%%%%%%%%
\subsection{Proof of C(2)}
As 
in the proof of B(2), we set 
\beq
X_{N, k}
&:=&
\frac {
Tr f_k (H_{\alpha, N}) - 
{\bf E}[ Tr f_k (H_{\alpha, N}) ] 
}
{
g_{6 \alpha}(N)
}, 
\quad
X_k
:=
N(0, \sigma_{C} (f_k)^2 )
\\
Y_N
&:=&
\frac {
Tr f (H_{\alpha, N}) - 
{\bf E}[ Tr f (H_{\alpha, N}) ] 
}
{
g_{6 \alpha}(N)
}, 
\quad
X
:=
N(0, \sigma_{C} (f)^2 )
\eeq
and we show that the three conditions (i), (ii), (iii) to apply Lemma 3.2 are satisfied.
(i)
has been done in 
\cite{B}.

(ii)
We shall show that, if 
$0 < \alpha \le 1/6$, 
the characteristic function 
$\varphi_k (t)$
of 
$N(0, \sigma_{C} (f_k)^2)$
satisfies
\beq
\varphi_k (t)
&=&
\exp
\left(
- \frac 12
\sigma_{C} (f_k)^2 
t^2
\right)
\to
\exp
\left(
- \frac 12
\sigma_{C} (f)^2 
t^2
\right).
\eeq
In fact,
we have
\beq
\sigma_{ C } (f_k)^2
&=&
\langle
f_k, C_{ C }^{(k)} f_k 
\rangle
=
\sum_{ 
\substack{ 
3 \le j \le k \\ j : odd
}
}
\sum_{ 
\substack{ 
3 \le j' \le k \\ j' : odd
}
}
c_j c_{j'}
M_{j, j'}
\eeq
where 
\beq
M_{j, j'}
&:=&
\lim_{N \to \infty}
Cov
\left(
\frac {
Tr P_j (H_{\alpha, N})
-
{\bf E}[ 
Tr P_j (H_{\alpha, N})
]
}
{
g_{6 \alpha} (N)
}, 
\frac {
Tr P_{j'} (H_{\alpha, N})
-
{\bf E}[ 
Tr P_j (H_{\alpha, N})
]
}
{
g_{6 \alpha} (N)
}
\right).
\eeq
Since 
\beq
| M_{j, j'} |
& \le &
\lim_{N \to \infty}
\left| Cov
\left(
\frac {
Tr P_j (H_{\alpha, N})
-
{\bf E}[ 
Tr P_j (H_{\alpha, N})
]
}
{
g_{6 \alpha} (N)
}, 
\frac {
Tr P_{j'} (H_{\alpha, N})
-
{\bf E}[ 
Tr P_j (H_{\alpha, N})
]
}
{
g_{6 \alpha} (N)
}
\right) 
\right|
\\
& \le &
\sum_{
\substack{
|\beta| \ge 3 \\ 
\iota(\beta) = 0
}
}
\sum_{
\substack{
|\gamma| \ge 3 \\ 
\iota(\gamma) = 0
}
}
p^j (\beta)
p^{j'} (\gamma)
\sum_{\ell \in {\bf Z}}
| 
Cov (X^{\beta}, X^{\gamma^{\ell}}) 
|
%\\
%
%& \le &
\le
\sum_{
\substack{
|\beta| \ge 3 \\ 
\iota(\beta) = 0
}
}
\sum_{
\substack{
|\gamma| \ge 3 \\ 
\iota(\gamma) = 0
}
}
p^j (\beta)
p^{j'} (\gamma)
j \cdot j' \cdot
4 C_X^{|\beta| + |\gamma|}
\\
& \le &
j (C_X+2)^j j' (C_X + 2)^{j'}
\eeq
we have
\beq
\sigma_{ C } (f_k)^2
& \le &
\sum_{ 
\substack{ 
3 \le j \le k \\ j : odd
}
}
\sum_{ 
\substack{ 
3 \le j' \le k \\ j' : odd
}
}
|c_j| \cdot |c_{j'}|
\cdot j \cdot j' \cdot
(C_X+2)^j  (C_X + 2)^{j'}
\eeq
which ensures the convergence of 
$\lim_{k \to \infty}
\sigma_{ C } (f_k)^2$.
\\
(iii)
By assumption, 
$c_j = 0$
for 
$j$ : even
so that 
${\cal C}_{i,j} (\beta) = \emptyset$, 
$\forall i$
for 
$| \beta | \notin 2 {\bf N} - 1$.
We compute 
\beq
&&
Var
\left(
\sum_{j \ge k+1	}
c_j
\left\{
\frac {
Tr (H_{\alpha,N}^j)
-
{\bf E}[ 
Tr (H_{\alpha,N}^j)
]
}
{
g_{6 \alpha} (N)
}
-
p^j (\delta)
\frac {
Tr (H_{\alpha,N})
-
{\bf E}[ 
Tr (H_{\alpha,N})
]
}
{
g_{6 \alpha} (N)
}
\right\}
\right)
\\
& = &
\frac {1}{g_{6 \alpha} (N)^2}
Var
\Biggl(
\sum_{j \ge k+1}
c_j
\Biggl\{
\sum_{|\beta|=1}
a_N^j (\beta)
( V^{\beta} - {\bf E}[ V^{\beta} ] )
+
\sum_{|\beta|\ge 3}
a_N^j (\beta)
( V^{\beta} - {\bf E}[ V^{\beta} ] )
\\
&& 
\hspace{8em}
- 
p^j (\delta)
\sum_{i=1}^N
\left(
V^{\delta^i} - {\bf E}[ V^{\delta^i} ] 
\right)
\Biggr\}
\Biggr)
\\
&=&
\frac {1}{g_{6 \alpha} (N)^2}
Var
\Biggl(
\sum_{j \ge k+1}
c_j
\Biggl\{
\sum_{i=1}^N
\left(
a_N^j (\delta^i) - p^j (\delta)
\right)
\left(
V^{\delta^i} - {\bf E}[ V^{\delta^i} ] 
\right)
+
\sum_{|\beta|\ge 3}
a_N^j (\beta)
( V^{\beta} - {\bf E}[ V^{\beta} ] )
\Biggr\}
\Biggr)
\\
& \le &
\frac {2}{g_{6 \alpha} (N)^2}
\Biggl\{
Var
\Biggl(
\sum_{j \ge k+1}
c_j
\sum_{|\beta|\ge 3}
a_N^j (\beta)
( V^{\beta} - {\bf E}[ V^{\beta} ] )
\Biggr)
\\
&& 
\hspace{5em}
+
Var
\Biggl(
\sum_{j \ge k+1}
c_j
\Biggl\{
\sum_{i=1}^N
\left(
a_N^j (\delta^i) - p^j (\delta)
\right)
\left(
V^{\delta^i} - {\bf E}[ V^{\delta^i} ] 
\right)
\Biggr)
\Biggr\}
\\
&=:&
2 (I + II).
\eeq
For 
$I$, 
\beq
I
&=&
{\bf E}
\left[
\left\{
\sum_{j \ge k+1	}
c_j
\sum_{ |\beta| \ge 3}
\frac {
a_N^j (\beta) 
\left(
V^{\beta} - {\bf E}[ V^{\beta} ] 
\right)
}
{
g_{6 \alpha} (N)
}
\right\}^2
\right]
\\
& \le &
\frac {1}{ g_{6 \alpha} (N)^2 }
\sum_{ j_1 \ge k+1 }
\sum_{j_2 \ge k+1} 
|c_{j_1}|\cdot  |c_{j_2}|
a_N^{j_1} (\beta)
a_N^{j_2} (\gamma)
|Cov (V^{\beta}, V^{\gamma})|
\\
& \le &
\frac {1}{ g_{6 \alpha} (N)^2 }
\sum_{ j_1 \ge k+1 }
\sum_{j_2 \ge k+1} 
|c_{j_1}|\cdot  |c_{j_2}|
\sum_{
\substack{ \iota(\beta)=0 \\ |\beta| \ge 3 }
}
\sum_{
\substack{ \iota(\gamma)=0 \\ |\gamma| \ge 3 }
}
p^{j_1} (\beta)
p^{j_2} (\gamma)
\sum_{\ell \in {\bf Z}}
| Cov (X^{\beta} X^{\gamma^{\ell}}) |
\\
&& \times
\sum_{ i\in I_{N, \ell} }
\prod_{n \in {\bf N}}
\frac {1}{
n^{\alpha (\beta_{n-i} + \gamma_{n - i - \ell})}
}
\\
& \le &
\frac {1}{ g_{6 \alpha} (N)^2 }
\sum_{ j_1 \ge k+1 }
\sum_{j_2 \ge k+1} 
|c_{j_1}|\cdot  |c_{j_2}|
\sum_{
\substack{ \iota(\beta)=0 \\ |\beta| \ge 3 }
}
\sum_{
\substack{ \iota(\gamma)=0 \\ |\gamma| \ge 3 }
}
p^{j_1} (\beta)
p^{j_2} (\gamma)
\cdot j_1 \cdot j_2 \cdot
4 C_X^{|\beta| + |\gamma|}
\cdot
\sum_{i=1}^N
\frac {1}{ i^{6 \alpha} }
\\
& \le &
4
\frac {
\sum_{i=1}^N
i^{- 6 \alpha}
}
{
g_{6 \alpha} (N)^2
}
\left(
\sum_{j \ge k+1}
|c_j| \cdot j \cdot
(C_X+2)^j
\right)^2.
\eeq
For 
$II$, 
\beq
II
&=&
\frac {1}{ g_{6 \alpha}(N)^2 }
Var
\left(
\sum_{j \ge k+1}
c_j
\sum_{i=1}^N
\left(
a_N^j (\delta^i) - p^j (\delta)
\right)
( V^{\delta^i} - {\bf E}[ V^{\delta^i} ] )
\right)
\\
&=&
\frac {1}{ g_{6 \alpha}(N)^2 }
\sum_{i=1}^N
\left\{
\sum_{j \ge k+1}
c_j
\left(
a_N^j (\delta^i) - p^j (\delta)
\right)
\right\}^2
\frac { {\bf E}[ X_i^2 ] }{i^{2 \alpha}}
\\
&=&
\frac {\eta^2}{ g_{6 \alpha}(N)^2 }
\sum_{j_1 \ge k+1}
\sum_{j_2 \ge k+1}
c_{j_1}
c_{j_2}
\sum_{i=1}^N
\left(
a_N^{j_1} (\delta^i) - p^{j_1} (\delta)
\right)
\left(
a_N^{j_2} (\delta^i) - p^{j_2} (\delta)
\right)
\frac {1}{i^{2 \alpha}}.
\eeq
Here we decompose 
\beq
\sum_{j_1 \ge k+1}
\sum_{j_2 \ge k+1}
&=&
\sum_{j_1=k+1}^{N}
\sum_{j_2=k+1}^{N}
+
\sum_{j_1=k+1}^{N}
\sum_{j_2=N+1}^{\infty}
+
\sum_{j_1=N+1}^{\infty}
\sum_{j_2=k+1}^{N}
+
\sum_{j_1=N+1}^{\infty}
\sum_{j_2=N+1}^{\infty}
\eeq
Since 
$a_N^j (\delta^i) = p^j (\delta)$
for 
$\frac j2 \le i \le N - \frac j2$, 
\beq
II
&=&
\frac {\eta^2}{g_{6 \alpha}(N)^2}
\sum_{j_1 = k+1}^N
\sum_{j_2 = k+1}^N
c_{j_1}
c_{j_2}
\left(
\sum_{i=1}^{
\min 
\left(
\frac {j_1-1}{2},\frac {j_2-1}{2} 
\right)
}
+
\sum_{
i = N - \min 
\left(
\frac {j_1-1}{2},\frac {j_2-1}{2} 
\right)
}^N
\right)
\\
&& 
\hspace{3em}
\cdot
\left(
a_N^{j_1} (\delta^i) - p^{j_1} (\delta)
\right)
\left(
a_N^{j_2} (\delta^i) - p^{j_2} (\delta)
\right)
\frac {1}{i^{2 \alpha}}
\\
&&
+
\frac {\eta^2}{ g_{6 \alpha}(N)^2 }
\left(
\sum_{j_1 = k+1}^N
\sum_{j_2 = N+1}^{\infty}
+
\sum_{j_1 = N+1}^{\infty}
\sum_{j_2 = k+1}^N
+
\sum_{j_1 = N+1}^{\infty}
\sum_{j_2 = N+1}^{\infty}
\right)
c_{j_1}
c_{j_2}
\\
&&
\hspace{3em}
\cdot
\sum_{i=1}^N
\left(
a_N^{j_1} (\delta^i) - p^{j_1} (\delta)
\right)
\left(
a_N^{j_2} (\delta^i) - p^{j_2} (\delta)
\right)
\frac {1}{i^{2 \alpha}}
\\
& \le &
\frac {\eta^2}{g_{6 \alpha}(N)^2}
\sum_{j_1 = k+1}^N
\sum_{j_2 = k+1}^N
|c_{j_1}|
|c_{j_2}|
\min ( j_1, j_2 )
\cdot
j_1 2^{j_1}
\cdot
j_2 2^{j_2}
\\
&&
+
\frac {\eta^2}{ g_{6 \alpha}(N)^2 }
\left(
\sum_{j_1 = k+1}^N
\sum_{j_2 = N+1}^{\infty}
+
\sum_{j_1 = N+1}^{\infty}
\sum_{j_2 = k+1}^N
+
\sum_{j_1 = N+1}^{\infty}
\sum_{j_2 = N+1}^{\infty}
\right)
|c_{j_1}|
|c_{j_2}|
\cdot
N
\cdot
j_1 2^{j_1}
\cdot
j_2 2^{j_2}.
\eeq
Here we used the following estimates. 
\beq
\quad
\sum_{i=1}^N
\frac {1}{i^{2 \alpha}}
\le N, 
\qquad
| a_N^j (\delta^i) - p^j (\delta) |
\le
p^j (\delta)
\le
j \cdot 2^j.
\eeq
We note that 
$N \le j_2$
in the 2nd term, 
$N \le j_1$
in the 3rd term, and 
$N \le j_1$
in the 4th term.
Therefore 
\beq
II
\le 
\frac {\eta^2}{g_{6 \alpha}(N)^2}
\left(
\sum_{j=k+1}^{\infty}
|c_j|
\cdot
j^2
\cdot
2^j
\right)^2.
\eeq
Plugging  
these estimates for 
$I, II$
above into Chebyshev's inequality, we have 
\beq
&&
{\bf P}
\left(
\left|
\sum_{j \ge k+1	}
c_j
\left(
\frac{
Tr H_{\alpha, N}^j
-
{\bf E}[
Tr H_{\alpha, N}^j
]
}
{
g_{6 \alpha} (N)
}
-
p^j (\delta)
\frac{
Tr H_{\alpha, N}
-
{\bf E}[
Tr H_{\alpha, N}
]
}
{
g_{6 \alpha} (N)
}
\right)
\right|
>
\epsilon
\right)
\\
& \le &
\frac {
2 (I+II)
}
{
\epsilon^2
}
\\
& \le &
\frac {2}{\epsilon^2}
\left[
\frac {
4 \sum_{i=1}^N i^{- 6 \alpha}
}
{
g_{6 \alpha} (N)^2
}
\left(
\sum_{j \ge k+1}
|c_j|
j (C_X+2)^j
\right)^2
+
\frac {\eta^2}{g_{6 \alpha}(N)^2}
\left(
\sum_{j=k+1}^{\infty}
|c_j|
\cdot
j^2
\cdot
2^j
\right)^2
\right]
\\
& \stackrel{N \to \infty}{\to}&
\frac {2 \cdot 4}{\epsilon^2}
\left(
\sum_{j \ge k+1}
|c_j|
j (C_X+2)^j
\right)^2
\stackrel{k \to \infty}{\to} 
0
\eeq
which yields the conclusion.
\QED
%

%
%%%%%%%%%%%%%%%%%%%%%%%%%%%%%%%%%%%%%%%%%%%%%%%%%%%%%%%%%%%%
\section{Proof of Theorem 2}
Take any 
${\bf \theta}
:=
(\theta_1, \cdots, \theta_d)
\in {\bf R}^d$ 
and let 
$\widetilde{F}_{\theta, N}
:=
{\bf \theta} \cdot {\bf F}_N$, 
$f_{\theta}
:=
\sum_{t=1}^d 
\theta_t f_t$, 
$\widetilde{F}_{\theta}
\sim
N(0, \sigma (f_{\theta})^2 )$.
By Theorem 1, we have 
\beq
\widetilde{F}_{\theta, N}
&:=&
\sum_{t=1}^d 
\theta_t F_{t, N}
=
\frac {1}{g_{\alpha/\alpha_c}(N)}
\left(
Tr (f_{\theta}(H_{\alpha, N}))
-
{\bf E}[ Tr (f_{\theta}(H_{\alpha, N}))] 
\right)
\stackrel{d}{\to} 
N(0, \sigma (f_{\theta})^2 )
\eeq
which shows 
$\widetilde{F}_{\theta, N}
\stackrel{d}{\to}
\widetilde{F}_{\theta}$. 
We denote by 
$\varphi_X (t)$
the characteristic function of the random variable 
$X$.
Since 
$\widetilde{F}_{\theta}
\sim
N(0, \sigma (f_{\theta})^2)$
and since 
$r(f) > C_X + 2$, 
we have 
\beq
\varphi_{ \widetilde{F}_{\theta} } (\xi)
&=&
\exp
\left[
- \frac 12
\sigma (f_{\theta})^2 \xi^2
\right]
=
\exp
\left[
- \frac 12
\xi^2
\left(
\sum_{t=1}^d
\theta_t
\sigma (f_t)
\right)^2
\right]
=
\varphi_{ {\bf \theta} \cdot {\bf F} }
(\xi)
\eeq
thus 
$\widetilde{F}_{\theta}
\stackrel{d}{=}
\theta \cdot {\bf F}$.
By the method of 
Cramer-Wald, this implies 
${\bf F}_N
\stackrel{d}{\to}
{\bf F}$, 
completing the proof of Theorem 2.
%
%
%%%%%%%%%%%%%%%%%%%%%%%%%%%%%%%%%%%%%%%%%%%%%%%%%%%%%%%%%%%%
\section{Proof of Theorem 3 : Polynomial case}
In this section
we prove 
Theorem 3 when 
$f$
is a polynomial of degree 
$k$.
General case 
will be considered in the next section.
Our goal in this section is : \\
%

%%%%%
{\bf Theorem 5.1 }
\begin{equation}
{\bf E}[ Tr (H_{\alpha,N}^k) ]
=
A_{k,1}N + A_{k, 0}
+
B^k (N)
+
\sum_{j=1}^k
C_{j, k}
\sum_{i=1}^N
\frac {1}{i^{\alpha j}}
+
D^k (N)
\label{5.1}
\end{equation}
where 
\beq
A_{k,1} 
&=& 
\left(
\begin{array}{c}
k \\ k/2
\end{array}
\right)
1 \left(
k : even
\right), 
\quad
A_{k,0}
= 
-
\sum_{{\bf y} \in {\cal R}^k}
(\max {\bf y}-\min {\bf y})
\\
{\cal R}^k 
&:=&
\left\{
{\bf y} = (y_0, \cdots, y_k) 
\in 
{\cal P}^k
\, \middle| \,
y_0 = y_k = 0, 
\;
{\bf y}
\mbox{ has no flat steps }
\right\}
\\
C_{j, k}
&:=&
\sum_{
\substack{
\beta \, : \, \iota(\beta) = 0 \\ |\beta| = j
}
}
p^k (\beta)
{\bf E}[ X^{\beta} ]
\\
B^k (N)
&:=&
\sum_{ \iota (\beta) \in [k, N-k]^c}
(a_{\beta} - p^k (\beta))
{\bf E}[ V_{\alpha}^{\beta} ]
\\
B^k (N)
&\stackrel{N \to \infty}{\to}& 
B^k, 
\quad
B^k 
:=
\sum_{ \iota (\beta) \in [1, k]}
(a_{\beta} - p^k (\beta))
{\bf E}[ V_{\alpha}^{\beta} ]
\\
D^k (N)
&=&
\sum_{j=1}^k
\sum_{
\substack{
\iota(\beta) = 0 \\ |\beta| = j
}
}
p^k (\beta)
{\bf E}[ X^{\beta} ]
\sum_{i=1}^N
\left(
\prod_{l \in {\bf N}}
\frac {1}{
l^{ \alpha \beta_{l-i} }
}
-
\frac {1}{i^{\alpha |\beta|}}
\right)
\\
D^k (N)
&\stackrel{N \to \infty}{\to}&
D^k, 
\quad
D^k
:=
\sum_{j=1}^k
\sum_{
\substack{
\iota(\beta) = 0 \\ |\beta| = j
}
}
p^k (\beta)
{\bf E}[ X^{\beta} ]
\sum_{i=1}^{\infty}
\left(
\prod_{l \in {\bf N}}
\frac {1}{
l^{ \alpha \beta_{l-i} }
}
-
\frac {1}{i^{\alpha |\beta|}}
\right)
%(- b (\beta))
\\
|D^k|
& \le &
\sum_{j=1}^k
\sum_{
\substack{
\iota(\beta) = 0 \\ |\beta| = j
}
}
p^k (\beta)
{\bf E}[ |X|^{\beta} ]
\sum_{i=1}^k
\frac {1}{i^{\alpha |\beta|}}
\eeq
%
%%%%%
%

(i)
$A^k=A_{k,1}N + A_{k, 0}$
is the contribution from the paths with no flat steps, which can be derived explicitly.
(ii)
$B^k (N)$
is a boundary effect. 
(iii)
$\sum_{j=1}^k
C_{j, k}
\sum_{i=1}^N
\dfrac {1}{i^{\alpha j}}
=
{\cal O}(
N^{1 - j\alpha}
)$
is the main term. 
The asymptotic behavior for 
$Tr (P(H_{\alpha,N}))$
for a polynomial
$P$
can be derived by summing up eq. 
(\ref{5.1}).
\begin{proof}
As in the proof of Theorem 1, 
let 
$a_{\beta}
:=
[V_{\alpha}^{\beta}]
\left(
Tr (H_{\alpha,N}^k)
\right)$
be the coefficient of 
$V^{\beta}$, 
$\beta \ne 0$
in 
$Tr (H_{\alpha,N}^k)$.
Then by Lemma 2.3, 
\beq
{\bf E}[
Tr H_{\alpha, N}^k
]
&=&
\sum_{ | \beta| \ge 1}
a_{\beta} 
{\bf E}[ V_{\alpha}^{\beta} ]
+
A^k (N)
=:
A^k (N)
+
B^k (N)
+
C^k (N)
\eeq
where 
$A^k (N)$
is the contribution of the paths with no flat steps which we write 
\beq
A^k (N)
&=&
A_{k,1}N + A_{k, 0}
%a_1^{(k)} N + a_0^{(k)}
\eeq
and 
\beq
B^k (N)
&:=&
\sum_{ \iota (\beta) \in [k, N-k]^c}
(a_{\beta} - p^k (\beta))
{\bf E}[ V_{\alpha}^{\beta} ]
\\
C^k (N)
&:=&
\sum_{ \iota(\beta) \in [1, N]}
p^k (\beta)
{\bf E}[ V_{\alpha}^{\beta} ]
\eeq
We compute these terms separately below. \\
(A)
We compute 
$A_k (N) = 
A_{k,1}N + A_{k, 0}$.
On the expansion of 
$H^k$, 
\beq
H^k
&=&
(V + U + D)^k
=
\sum_{
M \in {\cal M}
}
M
\eeq
we consider the terms 
$M = \prod_t M_t \in {\cal M}$
such that 
$M_t \ne V$
and 
$Tr M \ne 0$.
Since 
we must have 
$\sharp U = \sharp D$, 
in 
$\{ M_t \}$, 
$k$
must be even.
In that case, 
such terms are expressed as 
\beq
U^{u_1} D^{d_1} \cdots U^{u_j} D^{d_j}, 
\quad
u_1 + \cdots + u_j
=
d_1 + \cdots + d_j = k/2
\eeq
while the set of corresponding paths is 
\beq
{\cal R}^k 
&:=&
\left\{
{\bf y} = (y_0, \cdots, y_k) 
\in 
{\cal P}^k
\, \middle| \,
%y_0 = y_k = 0, 
\;
{\bf y}
\mbox{ has no flat steps }
\right\}
\eeq
By an explicit computation, 
we see that, in the corresponding 
$M \in {\cal M}$, 
$\max {\bf y}(M)$ 
components in the leftmost diagpnal part and 
$-\min {\bf y}(M)$
components in the rightleast diagonal part 
are all zero.
Thus for even $k$, 
\beq
Tr 
\left(
(U+D)^k
\right)
&=&
\sum_{{\bf y}(M)}
\left\{
N - 
(\max {\bf y}(M)-\min {\bf y}(M))
\right\}
\\
&=&
N
\left(
\begin{array}{c}
k \\ k/2
\end{array}
\right)
-
\sum_{{\bf y} \in {\cal R}^k}
(\max {\bf y}-\min {\bf y}).
\eeq
(B)
Since 
$V_{\alpha}$
is decaying, 
$B^k (N)$
is bounded and convergent : 
\beq
B^k (N)
\stackrel{N \to \infty}{\to} 
B^k, 
\quad
B^k 
:=
\sum_{ \iota (\beta) \in [1, k]}
(a_{\beta} - p^k (\beta))
{\bf E}[ V_{\alpha}^{\beta} ]
\eeq
(C)
Using 
%
%\beq
$
{\bf E}[ V_{\alpha}^{\beta} ] 
=
\prod_l
\left[
\dfrac {
X^{\beta_l}
}
{
l^{ \alpha \beta_l }
}
\right]
=
{\bf E}[ X^{\beta} ]
\prod_{l \in {\bf N}}
\dfrac {1}{
l^{ \alpha \beta_l }
}
$, 
we have 
\begin{equation}
C^k (N)
=
\sum_{j=1}^k
\sum_{i=1}^N
\sum_{
\substack{
\iota(\beta) = i \\ |\beta| = j
}
}
p^k (\beta)
{\bf E}[ X^{\beta} ]
\prod_{l \in {\bf N}}
\frac {1}{
l^{ \alpha \beta_l }
}.
\label{Ck}
\end{equation}
Since 
$\iota (\beta) = i$, 
$| \beta | = j$, 
$p^k (\beta) \ne 0$, 
$\beta_l > 0$
implies  
$l 
\in 
\left[
i, i + \frac k2 
\right]
\subset
[i, i + k ]$, 
and thus 
\beq
\frac {1}{
(i + k)^{\alpha |\beta|}
}
=
\prod_{l \in {\bf N}}
\frac {1}{
(i+k)^{ \alpha \beta_l }
}
\le
\prod_{l \in {\bf N}}
\frac {1}{
l^{ \alpha \beta_l }
}
\le
\prod_{l \in {\bf N}}
\frac {1}{
i^{ \alpha \beta_l }
}
=
\frac {1}{i^{\alpha |\beta|}}
\eeq
We replace 
$\prod_{l \in {\bf N}}
\dfrac {1}{
l^{ \alpha \beta_l }
}$
by 
$\dfrac {1}{i^{\alpha |\beta|}}$
in 
(\ref{Ck}), 
and denote the error by 
$D^k(N)$ : 
\beq
C^k (N)
&=:&
\sum_{j=1}^k
\sum_{i=1}^N
\sum_{
\substack{
\iota(\beta) = i \\ |\beta| = j
}
}
p^k (\beta)
{\bf E}[ X^{\beta} ]
\frac {1}{i^{\alpha |\beta|}}
+
D^k (N)
\eeq
By some change of variables, we have 
\beq
D^k (N)
&=&
\sum_{j=1}^k
\sum_{i=1}^N
\sum_{
\substack{
\iota(\beta) = 0 \\ |\beta| = j
}
}
p^k (\beta)
{\bf E}[ X^{\beta} ]
\left(
\prod_{l \in {\bf N}}
\frac {1}{
l^{ \alpha \beta_{l-i} }
}
-
\frac {1}{i^{\alpha |\beta|}}
\right)
=
\sum_{j=1}^k
\sum_{
\substack{
\iota(\beta) = 0 \\ |\beta| = j
}
}
p^k (\beta)
{\bf E}[ X^{\beta} ]
(- b_N (\beta))
\\
&&
where 
\quad
b_N (\beta)
:=
\sum_{i=1}^N
\left(
\frac {1}{i^{\alpha |\beta|}}
-
\prod_{l \in {\bf N}}
\frac {1}{
l^{ \alpha \beta_{l-i} }
}
\right)
\eeq
Since 
$b_N (\beta)$
is monotonically increasing and bounded,
\beq
\lim_{N \to \infty}
b_N (\beta) 
=
b(\beta)
:=
\sum_{i=1}^{\infty}
\left(
\frac {1}{i^{\alpha |\beta|}}
-
\prod_{l \in {\bf N}}
\frac {1}{
l^{ \alpha \beta_{l-i} }
}
\right).
\eeq
Therefore
\beq
D^k (N)
\stackrel{N \to \infty}{\to}
D^k, 
\quad
D^k
&:=&
\sum_{j=1}^k
\sum_{
\substack{
\iota(\beta) = 0 \\ |\beta| = j
}
}
p^k (\beta)
{\bf E}[ X^{\beta} ]
\sum_{i=1}^{\infty}
\left(
\prod_{l \in {\bf N}}
\frac {1}{
l^{ \alpha \beta_{l-i} }
}
-
\frac {1}{i^{\alpha |\beta|}}
\right)
%(- b (\beta))
\\
|D^k|
& \le &
\sum_{j=1}^k
\sum_{
\substack{
\iota(\beta) = 0 \\ |\beta| = j
}
}
p^k (\beta)
{\bf E}[ |X|^{\beta} ]
\sum_{i=1}^k
\frac {1}{i^{\alpha |\beta|}}
\eeq
On the other hand, 
\beq
C^k (N) - D^k (N)
&=&
\sum_{j=1}^k
C_{j, k}
\sum_{i=1}^N
\frac {1}{i^{\alpha j}}, 
\quad
C_{j, k}
:=
\sum_{
\substack{
\iota(\beta) = 0 \\ |\beta| = j
}
}
p^k (\beta)
{\bf E}[ X^{\beta} ]. 
\eeq
Here we note that
$C_{1,k} = 0$, 
since 
$\{ \beta 
\, | \,
\iota(\beta) = 0, 
|\beta| = 1
\}
=
\{ \delta \}$.
The proof of Theorem 5.1
is now complete.
\QED
\end{proof}
%
%

%
%%%%%%%%%%%%%%%%%%%%%%%%%%%%%%%%%%%%%%%%%%%%%%%%%%%%%%%%%%%%
\section{
Proof of Theorem 3 : general case}
In this section, 
we use Theorem 5.1 to finish the proof of Theorem 3.
To be more concrete, we show\\

{\bf Theorem 6.1}
\beq
&&
{\bf E}
[ Tr f(H) ] 
\\
&=&
\left(
\sum_l
c_l C_{0,l}
\right)
N
+
\left(
\sum_l
c_l C_{2,l}
\right)
S_2(N)
%N^{1 - 2 \alpha}
+
\cdots
+
\left(
\sum_l
c_l C_{m,l}
\right)
S_m (N)
%N^{1 - m \alpha}
+
C_N (f)
\\
&&
C_N (f)
:=
\left(
\sum_l 
c_l 
A_{l, 0}
\right)
+
\sum_{l \ge m+1}
c_l
\sum_{j=m+1}^l
C_{j, l} S_j(N)
+
\sum_{l}
c_l B^l (N)
+
\sum_{l}
c_l D^l (N)
\eeq
where 
\beq
\lim_{N \to \infty}
C_N (f)
&=&
\left(
\sum_l 
c_l 
A_{l, 0}
\right)
+
\sum_{l \ge m+1}
c_l
\sum_{j=m+1}^l
C_{j, l} S_j
+
\sum_{l}
c_l B^l
+
\sum_{l}
c_l D^l
\\
S_j
&:=&
\sum_{n = 1}^{\infty}
\frac {1}{
n^{j \alpha}
}, 
\quad
j \ge m+1
\\
B^k 
&:=&
\sum_{ \iota (\beta) \in [1, k]}
(a_{\beta} - p^k (\beta))
{\bf E}[ V_{\alpha}^{\beta} ]
\\
D^k
&:=&
\sum_{j=1}^k
\sum_{
\substack{
\iota(\beta) = 0 \\ |\beta| = j
}
}
p^k (\beta)
{\bf E}[ X^{\beta} ]
\sum_{i=1}^{\infty}
\left(
\prod_{l \in {\bf N}}
\frac {1}{
l^{ \alpha \beta_{l-i} }
}
-
\frac {1}{i^{\alpha |\beta|}}
\right)
\eeq
\begin{proof}
By Theorem 5.1, 
\begin{eqnarray}
{\bf E}
[ Tr H^k ] 
&=&
C_{0, k} N
+
C_{2, k} S_2(N)
+
\cdots
+
C_{m, k} S_m(N)
%\\
%
%&& \qquad
+
C_{m+1, k} S_{m+1}(N)
+
\cdots
+
C_{k, k} S_k(N)
\nonumber
\\
&&
+
B^k (N)
+
D^k (N)
+
A_{k, 0}
\label{trace}
%\quad\cdots (*)
\end{eqnarray}
where
$C_{0,k} = A_{k, 1}$, 
$S_{m+1}(N), \cdots, S_k(N)$
are bounded w.r.t. 
$N$.
If 
$k \le m$, 
we do not have terms of the form 
$C_{m+1, k} S_{m+1}
+
\cdots
+
C_{k, k} S_k$.
As 
we mentioned in the proof of Theorem 1, 
$f(H)
=
\sum_{l \ge 0}
a_l H^l$
is norm convergent, so that by Fubini theorem, 
\beq
{\bf E}[
Tr f(H)
]
&=&
\sum_{l \ge 0}
c_l
{\bf E}[
Tr (H^l)]. 
\eeq
Plugging it into 
(\ref{trace})
yields
\beq
&&
{\bf E}
[ Tr f(H) ] 
\\
&=&
\sum_l
c_l
{\bf E}[ Tr (H^l) ]
\\
&=&
\sum_l
c_l
\Bigl(
C_{0, l} N
+
A_{l, 0}
%a_0^{(l)}
+
C_{2, l} 
S_2(N)
%N^{1 - 2 \alpha}
+
\cdots
+
C_{m, l} 
S_m(N)
%N^{1 - m \alpha}
\\
&&
\qquad
+
C_{m+1, l} S_{m+1}(N)
+
\cdots
+
C_{l, l} S_l(N)
%\\
%
%&&
%\qquad
+
B^l (N)
+
D^l (N)
\Bigr)
\eeq
Here 
we would like to change the order of summation.
In order for that, 
it suffices to show that, 
\beq
&(i)&
\quad
\sum_{l \ge j}
c_l
C_{j, l}, 
\quad
(ii)
\quad
\sum_{l \ge m+1}
c_l
\sum_{j=m+1}^l
C_{j, l} S_j(N), 
\quad
(iii)
\quad
\sum_{l}
c_l B^l (N)
\\
&(iv)&
\quad
\sum_{l}
c_l D^l (N), 
\quad
(v)
\quad
\sum_l
c_l 
A_{l, 0}
\eeq
are absolutely convergent, and the quantities in 
(ii), (iii), (iv)
converge as 
$N \to \infty$.\\
(i)
By Lemma 2.4 (2), 
we have 
$|C_{j, l}| \le (C_X+2)^l$
so that since
$r(f) > C_X + 2$, 
(i)
is absolutely convergent. \\
%
%%%%%%%%%%%%%%%%%%%%%%%%%%%%%%%%%%%%
(ii)
That 
(ii)
is absolutely convergent is shown similarly as 
(i).
Since 
\beq
S(N)
&:=&
\max_{j \ge m+1}
S_j(N)
< \infty, 
\quad
\lim_{N \to \infty}
S_j (N)
=
S_j
:=
\sum_{n = 1}^{\infty}
\frac {1}{
n^{j \alpha}
}, 
\quad
j \ge m+1
\eeq
we have
\beq
\lim_{N \to \infty}
\sum_{l \ge m+1}
c_l
\sum_{j=m+1}^l
C_{j, l} S_j(N)
=
\sum_{l \ge m+1}
c_l
\sum_{j=m+1}^l
C_{j, l} S_j.
\eeq
%
%
%%%%%%%%%%%%%%%%%%%%%%%%%%%%%%%%%%%%
(iii)
Absolute convergence is similar as (i), (ii), and it is easy to see
\beq
\lim_{N \to \infty}
\sum_l
c_l B^l (N)
=
\sum_l
c_l B^l.
\eeq
(iv)
We have 
only to use the following estimate : 
\beq
|D^l (N)|
&\le&
\sum_{j=1}^l
\sum_{
\substack{
\iota(\beta) = 0 \\
|\beta| = j
}
}
p^l (\beta)
C_X^j
(Const.)
\max \{
1, 
l^{1 - \alpha |\beta|}
\}
\\
& \le &
(C_X + 2)^l 
(Const.)
\max \{
1, 
l^{1 - \alpha |\beta|}
\}.
\eeq
\\
(v)
By Stirling's formula, 
we have 
\beq
|A_{k, 0}|
&\le&
\sum_{ 
{\bf y} 
\in
{\cal R}_k
}
\left(
\max {\bf y} - \min {\bf y}
\right)
\le
k 
\left(
\begin{array}{c}
k \\ \frac k2
\end{array}
\right)
\le
(Const.)
\sqrt{k}
2^k
\eeq
so that 
$\sum_l c_l \cdot A_{l, 0}$
is absolutely convergent. 
The proof of Theorem 6.1
is now complete.
\QED
\end{proof}

\vspace*{1em}

%\noindent {\bf Acknowledgement }
%This work is partially supported by 
%Grant-in-Aid for Scientific Research (C) no.26400145.
This work is partially supported by 
JSPS KAKENHI Grant 
Number 20K03659(F.N.).

%%%%% REFERENCES %%%%%%%%%%%%%%%%%%%%%
%


\small
\begin{thebibliography}{99}

%
\bibitem{Bil}
Billingley, P., : 
Probability and Measure, 
3rd edition, 
Wiley series in probability and mathematical statistics, 
A Wiley-Interscience Publ. 

%
\bibitem{B}
Breuer, J., 
Grinshpon, Y., 
and
White M.J. : 
Spectral fluctuations for Schr\"odinger operators with a random decaying potential, 
Ann. Henri Poincar\'e
{\bf 22}(2021), 3763-3794.
%

%
\bibitem{DKS}
Delyon, F., Kunz, H., and Souillard, B. : 
From power pure point to continuous spectrum in disordered systems, 
Ann. Inst. H. Poincar\'e
{\bf 42}(1985), 283-309.

%
\bibitem{DE}
Dumitriu, I., and Edelman, A., 
Matrix models for beta ensembles, 
J. Math. Phys. 
{\bf 43}(2002), 5830-5847. 

%
\bibitem{D}
Durrett, R., :
Probability: Theory and Examples, 
5th edition, 
Cambridge Series in Statistical and Probabilistic Mathematics, 
Cambridge University Press. 


%
\bibitem{KLS}
Kiselev, A., Last, Y., and Simon, B. : 
Modified Pr\"ufer and EFGP Transforms and the Spectral Analysis of One-Dimensional Schr\"odinger Operators, 
Commun. Math. Phys. {\bf 194}(1997), 1 - 45.
%

%
\bibitem{KN1}
Kotani, S.,  and Nakano, F.,  
Level statistics for the one-dimensional Schroedinger operators with random decaying potential, 
Interdisciplinary Mathematical Sciences Vol. 17 (2014)
p.343-373.

%
\bibitem{KN2}
Kotani, S., and Nakano, F.,  
Poisson statistics for 1d Schr\"odinger operators with random decaying potentials, 
Electronic Journal of Probability {\bf 22}(2017), no.69, 1-31.

%
\bibitem{KVV}
Kritchevski, E., Valk\'o, B., Vir\'ag, B., : 
The scaling limit of the critical one-dimensional random 
Sdhr\"odinger operators, Commun. Math. Phys.{\bf 314}(2012), 775-806.
%

%
\bibitem{N1}
Nakano, F., 
Level statistics for one-dimensional Schr\"odinger operators and Gaussian beta ensemble, 
J. Stat. Phys.{\bf 156}(2014), 66-93.

%
\bibitem{N2}
Nakano, F.,
Fluctuation of density of states for 1d Schr\"odinger operators, 
J. Stat. Phys.{\bf 166}(2017):1393-1404. 


%
\bibitem{N3}
Nakano, F.,
Shape of eigenvectors for the decaying potential model, 
to appear in Annales Henri Poincar\'e. arXiv:2203.03125

%
\bibitem{RV}
Rifkind, B.,  Vir\'ag, B, : 
Eigenvectors of the 1-dimensional critical random Schr\"odinger operator, 
Geom. Funct. Anal. 
{\bf 28} (2018), 1394-1419. 
%




\end{thebibliography}
\end{document}